\documentclass[aps,prb,twocolumn,amsmath,amssymb,nofootinbib,superscriptaddress,floatfix,eqsecnum]{revtex4-1}

\usepackage{amsmath} 
\usepackage{amssymb} 
\usepackage{amsthm}
\usepackage[pdftex]{color} 
\usepackage{graphicx}
\usepackage{dcolumn} 
\usepackage{bm} 
\usepackage{longtable}
\usepackage{ulem} 
\begin{document}

\title{Wire deconstructionism of two-dimensional topological phases}

\author{Titus Neupert}
\affiliation{Princeton Center for Theoretical Science, Princeton University, Princeton,
New Jersey 08544, USA}

\author{Claudio Chamon} \affiliation{ Physics Department, Boston
  University, Boston, Massachusetts 02215, USA }
            
\author{Christopher Mudry} \affiliation{ Condensed Matter Theory
  Group, Paul Scherrer Institute, CH-5232 Villigen PSI, Switzerland}
  
  \author{Ronny Thomale} \affiliation{ Institute for Theoretical Physics, 
  University of W\"{u}rzburg, Am Hubland, D-97074 W\"{u}rzburg, Germany}

\date{\today}

\begin{abstract}
A scheme is proposed to construct integer and fractional 
topological quantum states of fermions in two spatial dimensions.
We devise models for such states by coupling wires of 
non-chiral Luttinger liquids of electrons, that are arranged 
in a periodic array. Which inter-wire couplings are allowed
is dictated by symmetry and the compatibility
criterion that they can simultaneously acquire a finite expectation value,
opening a spectral gap between the ground state(s) and all
excited states in the bulk.
First, with these criteria at hand, we reproduce the
tenfold classification table of integer topological insulators, 
where their stability against interactions becomes immediately transparent 
in the Luttinger liquid description.
Second, we construct an example of a strongly interacting fermionic topological phase of matter with short-range entanglement that lies outside of the tenfold classification. 
Third, we expand the table to long-range entangled topological phases
with intrinsic topological order and fractional excitations. 
\end{abstract}

\maketitle 

\section{Introduction}

The study of topological phases of matter is one of the most vibrant
directions of research in contemporary condensed matter physics. One
core accomplishment has been the theoretical
modeling and experimental discovery of two-dimensional
topological insulators.%
~\cite{Kane05a,Kane05b,Bernevig06,Koenig07} 
The integer quantum Hall effect (IQHE) was an early example of how states 
could be classified into distinct topological classes using an integer, 
the Chern number, to express the quantized Hall conductivity.%
~\cite{Klitzing80,Laughlin81,Thouless82}
In the IQHE, the number of delocalized
edge channels is directly tied to the quantized Hall conductivity
through the Chern number. More recently, it has been
found that the symmetry under reversal of time
acts as a protective symmetry for edge modes in (bulk)
insulators with strong spin-orbit interactions in two and
three dimensions,~\cite{Kane05a,Kane07} and that these systems are
characterized by a $\mathbb{Z}^{\,}_{2}$ topological invariant.

The discovery of $\mathbb{Z}^{\,}_{2}$ topological insulators has triggered
 a search for a classification of phases of 
fermionic matter that are distinct by some topological attribute. 
For non-interacting electrons, a complete classification, the tenfold way, 
has been accomplished in arbitrary dimensions.%
~\cite{Schnyder08,Kitaev09} 
In this scheme, three discrete symmetries that act locally in position space
-- time-reversal symmetry (TRS), 
particle-hole symmetry
(PHS), 
and chiral or sublattice symmetry (SLS) -- 
play a central role when defining the quantum numbers that
identify the topological insulating fermionic phases of matter
within one of the ten symmetry classes 
(see columns 1-3 from Table~\ref{table: main table}).

The tenfold way is believed to be robust to a perturbative treatment
of short-ranged electron-electron interactions for the following
reasons.  First, the unperturbed ground state in the clean limit and
in a closed geometry is non-degenerate and given by the filled bands 
of a band insulator. The band gap provides a small expansion parameter, namely
the ratio of the characteristic interacting energy scale to the band
gap. Second, the quantized topological invariant that characterizes
the filled bands, provided its definition and topological character
survives the presence of electron-electron interactions 
as is the case for the symmetry class A in two spatial dimensions,
cannot change in a perturbative treatment of 
short-range electron-electron interactions.%
~\cite{Gurarie11} 

On the other hand, the fate of the tenfold way when electron-electron
interactions are strong is rather subtle.%
~\cite{Fidkowski10,Gurarie11,Manmana12,Wang12} For example, short-range
interactions can drive the system through a topological phase
transition at which the energy gap closes,%
~\cite{Raghu08, Budich12}
or they may spontaneously break a defining symmetry of the topological
phase. Even when short-range interactions neither spontaneously break
the symmetries nor close the gap, it may be that two phases from the
non-interacting tenfold way cease to be distinguishable in the
presence of interactions. In fact, it was shown for the symmetry class
BDI in one dimension by Fidkowski and Kitaev that the non-interacting
$\mathbb{Z}$ classification was too fine in that it must be replaced
by a $\mathbb{Z}^{\,}_{8}$ classification when generic short-range
interactions are allowed. How to construct a counterpart
to the tenfold way for interacting fermion (and boson) systems has
thus attracted a lot of interest.%
~\cite{Gu09,Pollmann10,Cirac11,Chen11a,Chen11b,Fidkowski11,Turner11,Gu12,Lu12,Chen13,Lu13}

The fractional quantum Hall effect (FQHE) is the paradigm for a
situation by which interactions select topologically ordered ground states
of a very different kind than the non-degenerate ground states from
the tenfold way.  On a closed two-dimensional manifold of genus $g$, 
interactions can stabilize incompressible many-body ground states 
with a $g$-dependent degeneracy. 
Excited states in the bulk must then carry fractional
quantum numbers (see Ref.~\onlinecite{Wen91} and references therein).
Such phases of matter, that follow the FQHE paradigm,
appear in the literature under different names: fractional topological
insulators, long-range entangled phases, topologically ordered phases,
or symmetry enriched topological phases. In this paper we use
the terminology long-range entangled (LRE) phase for all phases 
with nontrivial $g$-dependent ground state degeneracy.  All other phases, 
i.e., those that follow the IQHE paradigm, are
called short-range entangled (SRE) phases.
(In doing so, we follow the terminology of Ref.~\onlinecite{Lu12}, 
that differs slightly from the one used in Ref.~\onlinecite{Chen11a}. 
The latter counts all chiral phases irrespective of their ground state 
degeneracy as LRE.)

While there are nontrivial SRE and LRE phases in the absence of any symmetry 
constraint,
many SRE and LRE phases are defined by some protecting symmetry they obey. 
If this protecting symmetry is broken, the topological attribute of the phase 
is not well defined any more.
However, there is a sense in which LRE phases are more robust
than SRE phases against a weak breaking of the defining symmetry.
The topological attributes of LRE phases are not confined to the
boundary in space between two distinct topological realizations of
these phases, as they are for SRE phases. They also characterize 
intrinsic bulk properties such as the existence of gapped deconfined 
fractionalized excitations. 
Hence, whereas gapless edge states are gapped by any breaking of the
defining symmetry, topological bulk properties are robust to a weak breaking
of the defining symmetry as long as the characteristic energy scale
for this symmetry breaking is small compared to the bulk gap
in the LRE phase, for a small breaking of the protecting
symmetry does not wipe out the gapped deconfined fractionalized 
bulk excitations.

The purpose of this paper is to implement a classification scheme for
interacting electronic systems in two spatial dimensions that treats 
SRE and LRE phases on equal footing. To this end, we use a coupled
wire construction for each of the symmetry classes from the tenfold
way. This approach has been pioneered in Refs.%
~\onlinecite{Yakovenko91} and \onlinecite{Lee94}
for the IQHE and in Refs.%
~\onlinecite{Kane02} 
and 
\onlinecite{Teo14}
for the FQHE
(see also related work in Refs.%
~\onlinecite{Sondhi01,Klinovaja13b,Klinovaja13a,gangof11,Seroussi14,PhysRevX.4.031009}).

To begin with, non-chiral Luttinger liquids are placed in a periodic
array of coupled wires.  
In doing so, forward-scattering two-body interactions 
are naturally accounted for within each wire. 
We then assume that the back-scattering (i.e., tunneling) within a given wire or
between neighboring wires are the dominant energy scales.
Imposing symmetries constrains these allowed
tunnelings.  
Whether a given arrangement of
tunnelings truly gaps out all bulk modes, except for some ungapped edge
states on the first and last wire, is verified with the help of a
condition that applies to the limit of strong tunneling.
We name this condition the Haldane criterion, as it was introduced by Haldane
in his study of the stability of non-maximally chiral edge states in
the quantum Hall effect.%
~\cite{Haldane95} We show that, for a proper choice of the tunnelings, 
all bulk modes are gapped.  Moreover, in
five out of the ten symmetry classes of the tenfold way, there remain
gapless edge states in agreement with the tenfold way.  It is the
character of the tunnelings that
determines if this wire construction selects a SRE or a LRE phase.
Hence, this construction, predicated as it is on the strong tunneling
limit, generalizes the tenfold way for SRE phases to LRE phases.  It
thereby delivers LRE phases that have not yet appeared in the
literature before. Evidently, this edge-centered classification
scheme does not distinguish between LRE phases of matter that do not
carry protected gapless edge modes at their interfaces. For example,
some fractional, time-reversal-symmetric, incompressible and topological phases
of matter can have fractionalized excitations in the bulk, 
while not supporting protected gapless modes at their boundaries.%
~\cite{Wen91a,Sachdev91,Mudry94}

Stated in a slightly more constructive way, we can think of our
approach as (1) fixing, in a first step, a given desired edge theory
at the boundary, and (2) continue, in a second step, by asking whether
such an edge can be consistently defined with a set of
symmetry-allowed periodic tunneling terms between wires which manage
to gap out all other modes. Alluding to a related strategy in
philosophy, this is what we call wire deconstructionism of
topological phases.

The paper is organized as follows.  We define the array of Luttinger
liquids in Sec.~\ref{sec: Definitions}.  The Haldane criterion, which
plays an essential role for the stability analysis of the edge theory,
is reviewed in Sec.~\ref{sec: Conditions for a spectral gap}.  All
five SRE entries of Table~\ref{table: main table} are derived in
Sec.~\ref{sec: Reproducing the tenfold way}, while all five LRE
entries of Table~\ref{table: main table} are derived in 
Sec.~\ref{sec: Fractionalized phases}.  We conclude with Sec.%
~\ref{sec: Discussion}, 
where we allude to the generalization
of our approach to additional symmetries, bosonic systems, and
higher spatial dimensions.

\begin{figure}[t]
\begin{center}
\includegraphics[width=84mm, page=1]{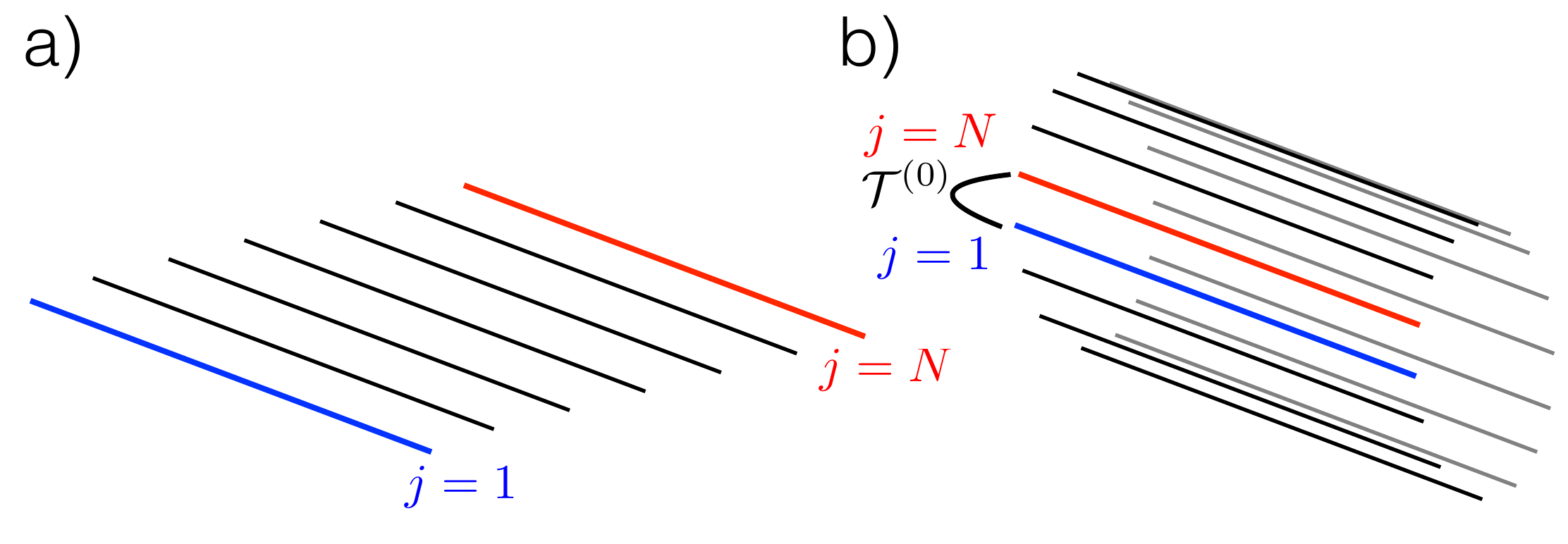}
\caption{(Color online) 
The boundary conditions determine whether a topological phase 
has protected gapless modes or not.
(a)
With open boundary conditions, gapless modes exist near the wires $j=1$ 
and $j=N$, 
the scattering between them is forbidden by imposing locality in the limit 
$N\to\infty$.
(b)
Periodic boundary conditions allow the scattering vector $\mathcal{T}^{(0)}$ 
that gaps modes which were protected by locality before.  
        }
\label{fig: BoundaryConditions}
\end{center}
\end{figure}

\begin{table*}[t]
\caption{(Color online) 
Realization of a two-dimensional array of
quantum wires in each symmetry class of the tenfold way.
For each of the symmetry classes A, AII, D, DIII, and C,
the ground state supports propagating gapless edge modes localized
on the first and last wire that are immune to local and
symmetry-preserving perturbations. The first column labels the
symmetry classes according to the Cartan classification of symmetric
spaces. The second column dictates if the operations for reversal of
time 
($\widehat{\Theta}$ with the single-particle representation 
$\Theta$), 
exchange of particles and holes
($\widehat{\Pi}$ with the single-particle representation 
$\Pi$), 
and reversal of chirality 
($\widehat{C}$ with the single-particle representation 
$C$)
are the generators of symmetries 
with their single-particle representations
squaring to $+1$,
$-1$, or are not present in which case the entry $0$ is used. 
(See the footnote~\onlinecite{footnotechiraltrsf} for a definition of
$\widehat{C}$.) 
The third column is the set to which the topological index from
the tenfold way, defined as it is in the non-interacting limit, belongs to.
The fourth column is a pictorial representation of the interactions 
(a set of tunnelings vectors $T$)
for the two-dimensional array of quantum wires that delivers short-range
entangled (SRE) gapless edge states. A wire is represented by
a colored box with the minimum number of channels compatible with the
symmetry class. Each channel in a wire is either a right mover
($\otimes$) or a left mover ($\odot$) that may or may not carry a
spin quantum number ($\uparrow,\downarrow$) or a particle (yellow
color) or hole (black color) attribute. The lines describe
tunneling processes within a wire or between consecutive wires 
in the array that are of one-body type 
when they do not carry an arrow
or of strictly many-body type when they carry an arrow.
Arrows point toward the sites on which creation operators act 
and away from the sites on which annihilation operators act.
For example in the symmetry class A, the single line connecting two
consecutive wires in the SRE column represents a one-body backward scattering
by which left and right movers belonging to consecutive wires are
coupled. The lines have been omitted for the fifth (LRE) column, 
only the tunneling vectors are specified.
       }
\begin{center}
\begin{tabular}{| l | c c c |  c | l | l |}
\hline
$\vphantom{\Bigg[}$
&$\Theta^{2}$ &$\Pi^{2}$&$C^{2}$& &
Short-range entangled (SRE) topological phase
&
Long-range entangled (LRE) topological phase
\\
\hline
\hline
$\vphantom{\Bigg[}$
A
&$0$ &$0$&$0$&$\mathbb{Z}^{\ }$&
\begin{tabular}{c}
\includegraphics[scale=0.35,page=1]{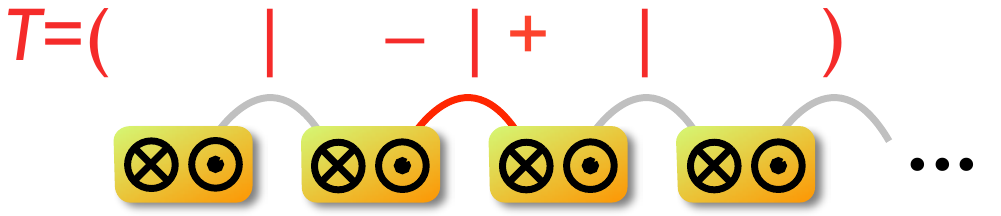}
\end{tabular}
&
\begin{tabular}{c}
\includegraphics[scale=0.35,page=2]{TableFigs.pdf}
\end{tabular}
\\
\hline
$\vphantom{\Bigg[}$
AIII
&$0$ &$0$&$+$& & NONE &\\
\hline
\hline
$\vphantom{\Bigg[}$
AII
&$-$ &$0$&$0$&$\mathbb{Z}^{\ }_{2}$&
\begin{tabular}{c}
\includegraphics[scale=0.35,page=3]{TableFigs.pdf}
\end{tabular}
&
\begin{tabular}{c}
\includegraphics[scale=0.35,page=4]{TableFigs.pdf}
\end{tabular}
\\
\hline
$\vphantom{\Bigg[}$
DIII
&$-$ &$+$&$+$&$\mathbb{Z}^{\ }_{2}$&
\begin{tabular}{c}
\includegraphics[scale=0.35,page=7]{TableFigs.pdf}
\end{tabular}
&
\begin{tabular}{c}
\includegraphics[scale=0.35,page=8]{TableFigs.pdf}
\end{tabular}
\\
$\vphantom{\Bigg[}$
D
&$0$ &$+$&$0$&$\mathbb{Z}^{\ }$&
\begin{tabular}{c}
\includegraphics[scale=0.35,page=5]{TableFigs.pdf}
\end{tabular}
&
\begin{tabular}{c}
\includegraphics[scale=0.35,page=6]{TableFigs.pdf}
\end{tabular}
\\
$\vphantom{\Bigg[}$
BDI
&$+$ &$+$&$+$& & NONE &\\
\hline
$\vphantom{\Bigg[}$
AI
&$+$ &$0$&$0$& & NONE &\\
\hline
$\vphantom{\Bigg[}$
CI
&$+$ &$-$&$+$& & NONE &\\
$\vphantom{\Bigg[}$
C
&$0$ &$-$&$0$&$\mathbb{Z}^{\ }$&
\begin{tabular}{c}
\includegraphics[scale=0.35,page=9]{TableFigs.pdf}
\end{tabular}
&
\begin{tabular}{c}
\includegraphics[scale=0.35,page=10]{TableFigs.pdf}
\end{tabular}
\\
CII
&$-$ &$-$&$+$& & NONE &\\
\hline
\end{tabular}
\end{center}
\label{table: main table}
\end{table*}%

\section{Definitions}
\label{sec: Definitions}

We consider an array of $N$ parallel wires that stretch along the $x$
direction of the two-dimensional embedding Euclidean space
(see Fig.~\ref{fig: BoundaryConditions}).
We label a
wire by the Latin letter $i=1,\cdots,N$. Each wire supports fermions
that carry an even integer number $M$ of internal degrees of freedom
that discriminate between left- and right-movers, the projection along
the spin-$1/2$ quantization axis, and particle-hole quantum numbers,
among others (e.g., flavors). We label these internal degrees of
freedom by the Greek letter $\gamma=1,\cdots,M$. We combine those two
indices in a collective index $\mathsf{a}\equiv
(i,\gamma)$. Correspondingly, we introduce the $M\times N$ pairs of
creation $\hat{\psi}^{\dag}_{\mathsf{a}}(x)$ and annihilation
$\hat{\psi}^{\,}_{\mathsf{a}}(x)$ field operators obeying the
fermionic equal-time algebra
\begin{subequations}
\label{eq: def hat psi's}
\begin{equation}
\left\{
\hat{\psi}^{\,}_{\mathsf{a}}(x),
\hat{\psi}^{\dag}_{\mathsf{a}'}(x')
\right\}=
\delta^{\,}_{\mathsf{a},\mathsf{a}'}\,
\delta(x-x')
\label{eq: def hat psi's a}
\end{equation}
with all other anticommutators vanishing and
the collective labels $\mathsf{a},\mathsf{a}'=1,\cdots,M\times N$.
The notation
\begin{equation}
\widehat{\Psi}^{\dag}(x)\equiv
\begin{pmatrix}
\hat{\psi}^{\dag}_{1}(x)&\cdots&\hat{\psi}^{\dag}_{MN}(x)
\end{pmatrix},
\quad
\widehat{\Psi}(x)\equiv
\begin{pmatrix}
\hat{\psi}^{\,}_{1}(x)\\ \vdots\\ \hat{\psi}^{\,}_{MN}(x)
\end{pmatrix},
\label{eq: def hat psi's b}
\end{equation}
\end{subequations}
is used for the operator-valued row ($\widehat{\Psi}^{\dag}$) 
and column ($\widehat{\Psi}^{\,}$) vector fields.
We assume that the many-body quantum dynamics of the fermions supported by 
this array of wires is governed by the Hamiltonian
$\hat{H}$, whereby interactions within each wire
are dominant over interactions between wires so that 
we may represent $\hat{H}$ as $N$ coupled Luttinger liquids,
each one of which is composed of $M$ interacting fermionic channels.

By assumption, we may thus bosonize the $M\times N$ 
fermionic channels making up the array. To this end,
we follow Ref.~\onlinecite{Neupert11}.
Within Abelian bosonization, this is done by postulating first 
the $MN\times MN$ matrix 
\begin{subequations}
\label{eq: def mathcal K}
\begin{equation}
\mathcal{K}\equiv
\left(\mathcal{K}^{\,}_{\mathsf{a}\mathsf{a}'}\right)
\label{eq: def mathcal K a}
\end{equation}
to be symmetric with integer-valued entries.
Because we are after an array of identical wires,
each of which having its quantum dynamics governed by
that of a Luttinger liquid, it is
natural to assume that $\mathcal{K}$ is reducible,
\begin{equation}
\mathcal{K}^{\,}_{\mathsf{a}\mathsf{a}'}=
\delta^{\,}_{ii'}\,
K^{\,}_{\gamma\gamma'},
\quad
i,i'=1,\cdots,N,
\quad
\gamma,\gamma'=1,\cdots,M.
\label{eq: def mathcal K c}
\end{equation}
\end{subequations}
A second $MN\times MN$ matrix is then defined by
\begin{subequations}
\label{eq: def mathcal L}
\begin{equation}
\mathcal{L}\equiv
\left(\mathcal{L}^{\,}_{\mathsf{a}\mathsf{a}'}\right)
\label{eq: def mathcal L a}
\end{equation}
where
\begin{equation}
\mathcal{L}^{\,}_{\mathsf{a}\mathsf{a}'}:=
\mathrm{sgn}(\mathsf{a}-\mathsf{a}')
\left(
\mathcal{K}^{\,}_{\mathsf{a}\mathsf{a}'}
+
1
\right).
\label{eq: def mathcal L b}
\end{equation}
\end{subequations}
Third, one verifies that,
for any pair $\mathsf{a},\mathsf{a}'=1,\cdots,MN$,
the Hermitian fields
$\hat{\phi}^{\,}_{\mathsf{a}}$ 
and
$\hat{\phi}^{\,}_{\mathsf{a}'}$, 
defined by the Mandelstam formula
\begin{subequations}
\label{eq: def hat phi's}
\begin{equation}
\hat{\psi}^{\,}_{\mathsf{a}}(x)\equiv\;\;
:
\exp
\left(
+\mathrm{i}
\mathcal{K}^{\,}_{\mathsf{a}\mathsf{a}'}\,
\hat{\phi}^{\,}_{\mathsf{a}'}(x)
\right)
:
\label{eq: def hat phi's a}
\end{equation}
as they are,
obey the bosonic equal-time algebra
\begin{equation}
\left[
\hat{\phi}^{\,}_{\mathsf{a} }(x ),
\hat{\phi}^{\,}_{\mathsf{a}'}(x')
\right]=
-\mathrm{i}\pi
\left(
\mathcal{K}^{-1}_{\mathsf{a}\mathsf{a}'}\,
\mathrm{sgn}(x-x')
+
\mathcal{K}^{-1}_{\mathsf{a}\mathsf{b}}\,
\mathcal{L}^{\,}_{\mathsf{b}\mathsf{c}}\,
\mathcal{K}^{-1}_{\mathsf{c}\mathsf{a}'}
\right).
\label{eq: def hat phi's b}
\end{equation}
Here, the notation $:(\cdots):$ stands for normal ordering of the
argument $(\cdots)$ and the summation convention over repeated indices
is implied. In line with Eq.~(\ref{eq: def hat psi's b}), we use the
notation
\begin{equation}
\widehat{\Phi}^{\mathsf{T}}(x)\equiv
\begin{pmatrix}
\hat{\phi}^{\,}_{1}(x)&\cdots&\hat{\phi}^{\,}_{MN}(x)
\end{pmatrix},
\quad
\widehat{\Phi}(x)\equiv
\begin{pmatrix}
\hat{\phi}^{\,}_{1}(x)\\ \vdots\\ \hat{\phi}^{\,}_{MN}(x)
\end{pmatrix},
\label{eq: def hat phi's c}
\end{equation}
for the operator-valued row 
($\widehat{\Phi}^{\mathsf{T}}$)
and column 
($\widehat{\Phi}$)
vector fields.
Periodic boundary conditions along the $x$ direction
parallel to the wires are imposed by demanding that
\begin{equation}
\mathcal{K}\,\widehat{\Phi}(x+L)=
\mathcal{K}\,\widehat{\Phi}(x)
+
2\pi\,\mathcal{N},
\qquad
\mathcal{N}\in\mathbb{Z}^{MN}.
\end{equation}
\end{subequations}

Equipped with 
Eqs.~(\ref{eq: def mathcal K})--(\ref{eq: def hat phi's}),
we decompose additively the many-body Hamiltonian $\hat{H}$ 
for the $MN$ interacting fermions propagating on the array of wires into 
\begin{subequations}
\label{eq: hat H bosonized}
\begin{equation}
\hat{H}=
\hat{H}^{\,}_{\mathcal{V}}
+
\hat{H}^{\,}_{\{\mathcal{T}\}}.
\label{eq: hat H bosonized a}
\end{equation}
Hamiltonian
\begin{equation}
\hat{H}^{\,}_{\mathcal{V}}:=
\int\mathrm{d}x\,
\left(\partial^{\,}_{x}\widehat{\Phi}^{\mathsf{T}}\right)(x)\;
\mathcal{V}\;
\left(\partial^{\,}_{x}\widehat{\Phi}\right)(x),
\label{eq: hat H bosonized b}
\end{equation}
even though quadratic in the bosonic field,
encodes both local one-body terms as well as
contact many-body interactions between
the $M$ fermionic channels in any given wire from the array 
through the block-diagonal, real-valued, and  symmetric $MN\times MN$ matrix
\begin{equation}
\label{eq: def mathcal V}
\mathcal{V}:=
\left(
\mathcal{V}^{\,}_{\mathsf{a}\mathsf{a}'}
\right)\equiv
\left(
\mathcal{V}^{\,}_{(i,\gamma)(i',\gamma')}
\right)=
\openone^{\,}_{N}\otimes
(V^{\,}_{\gamma\gamma'}).
\end{equation}
Hamiltonian
\begin{align}
\hat{H}^{\,}_{\{\mathcal{T}\}}:=&\,
\int\mathrm{d}x\,
\sum_{\mathcal{T}}
\frac{h^{\,}_{\mathcal{T}}(x)}{2}\,
\left(
e^{+\mathrm{i}\alpha^{\,}_{\mathcal{T}}(x)}
\prod_{\mathsf{a}=1}^{MN}
\hat{\psi}^{\mathcal{T}^{\,}_{\mathsf{a}}}_{\mathsf{a}}(x)
+
\mathrm{H.c.}
\right)
\nonumber\\
=&\,
\int\mathrm{d}x\,
\sum_{\mathcal{T}}
h^{\,}_{\mathcal{T}}(x)\,
\cos
\left(
\mathcal{T}^{\mathsf{T}}\,
\mathcal{K}\,
\widehat{\Phi}(x)
+
\alpha^{\,}_{\mathcal{T}}(x)
\right)
\label{eq: hat H bosonized c}
\end{align}
is not quadratic in the bosonic fields. With the understanding that
the operator-multiplication of identical fermion fields at the same
point $x$ along the wire requires point splitting, and with the
short-hand notation
$\hat{\psi}^{-1}_{\mathsf{a}}(x)\equiv\hat{\psi}^{\dag}_{\mathsf{a}}(x)$,
we interpret $\hat{H}^{\,}_{\{\mathcal{T}\}}$ as (possibly many-body)
tunnelings between the fermionic channels. Here, we introduced the
set $\{\mathcal{T}\}$ comprised of \textit{all} integer-valued
tunneling vectors
\begin{equation}
\mathcal{T}\equiv
\left(\mathcal{T}^{\,}_{\mathsf{a}}\right)
\label{eq: def mathcal T a}
\end{equation}
obeying the condition
\begin{equation}
\sum_{\mathsf{a}=1}^{MN}
\mathcal{T}^{\,}_{\mathsf{a}}=
\begin{cases}
0\hbox{ mod }2,
&
\hbox{ for D, DIII, C, and CI,}
\\
&
\\
0,
&
\hbox{ otherwise,}
\end{cases} 
\label{eq: def mathcal T b}
\end{equation}
and we assigned to each $\mathcal{T}$ from the set $\{\mathcal{T}\}$
the real-valued functions
\begin{equation}
h^{\,}_{\mathcal{T}}(x)=
h^{* }_{\mathcal{T}}(x)\geq0
\label{eq: def mathcal T c}
\end{equation}
and
\begin{equation}
\alpha^{\,}_{\mathcal{T}}(x)=
\alpha^{* }_{\mathcal{T}}(x).
\label{eq: def mathcal T d}
\end{equation}
\end{subequations}
The condition~(\ref{eq: def mathcal T b}) ensures that
these tunneling events preserve the parity of the total fermion number
for the superconducting symmetry classes
(symmetry classes D, DIII, C, and CI in Table \ref{table: main table}), 
while they preserve the total fermion number for the non-superconducting 
symmetry classes (symmetry classes A, AIII, AI, AII, BDI, and CII in Table 
\ref{table: main table}).
We emphasize that the integer
\begin{equation}
q:= \sum_{\mathsf{a}=1}^{MN}
\frac{|\mathcal{T}^{\,}_{\mathsf{a}}|}{2}
\label{eq:q-def}
\end{equation}
dictates that $\mathcal{T}$ encodes a $q$-body interaction
in the fermion representation.

\section{Strategy for constructing topological phases}
\label{sec: big strategy section}

Our strategy consists in choosing the many-body Hamiltonian 
$\hat{H}=\hat{H}^{\,}_{\mathcal{V}}+\hat{H}^{\,}_{\{\mathcal{T}\}}$
defined in Eq.~(\ref{eq: hat H bosonized})
so that
(i) it belongs to any one of the ten symmetry classes from 
the tenfold way (with the action of symmetries defined in Sec.~\ref{sec: representation of symmetries})
and (ii) all excitations in the bulk are gapped by a \textit{specific}
choice of the tunneling vectors $\{\mathcal{T}\}$ entering $\hat{H}^{\,}_{\{\mathcal{T}\}}$
(with the condition for a spectral gap given in Sec.~\ref{sec: Conditions for a spectral gap})
The energy scales in
$\hat{H}^{\,}_{\{\mathcal{T}\}}$ 
are assumed sufficiently large compared to those in $\hat{H}^{\,}_{\mathcal{V}}$ 
so that it is
$\hat{H}^{\,}_{\mathcal{V}}$
that may be thought of as a perturbation of 
$\hat{H}^{\,}_{\{\mathcal{T}\}}$ and not the converse. 

We anticipate that for five of the ten symmetry classes there can be 
protected gapless edge states because of locality and symmetry.
Step (ii) for each of the five symmetry classes 
supporting gapless edge states
is represented pictorially
as is shown in the fourth column of Table~\ref{table: main table}.
In each symmetry class, topologically trivial states that do not
support protected gapless edge states
in the tenfold classification can be constructed by gapping all 
states in each individual wire from the array.

\subsection{Representation of symmetries}
\label{sec: representation of symmetries}

The classification is based on the 
presence or the absence of the TRS 
and the PHS that 
are represented by the antiunitary many-body operator $\widehat{\Theta}$ 
and the unitary many-body operator $\widehat{\Pi}$, respectively.
Each of $\widehat{\Theta}$ and $\widehat{\Pi}$
can exist in two varieties such that their single-particle
representations $\Theta$ and $\Pi$
square to the identity operator
up to the multiplicative factor $\pm 1$,
\begin{equation}
\Theta^{2}=\pm 1,
\qquad
\Pi^{2}=\pm 1,
\label{eq: widehat Theta and widehat Phi square to pm1}
\end{equation} 
respectively. By assumption,
the set of all degrees of freedom in each given wire 
is invariant under the actions of $\widehat{\Theta}$ and $\widehat{\Pi}$.
If so, we can represent the actions of
$\widehat{\Theta}$ and $\widehat{\Pi}$
on the fermionic fields in two steps.
First, we introduce two $M\times M$-dimensional
matrix representations $P^{\,}_{\Theta}$ and $P^{\,}_{\Pi}$ 
of the permutation group of $M$ elements,
which we combine into the block-diagonal $MN\times MN$ 
real-valued and orthogonal matrices
\begin{subequations}
\label{eq: def reps widehat Theta and widehat Pi}
\begin{equation}
\mathcal{P}^{\,}_{\Theta}:=\openone^{\,}_{N}\otimes P^{\,}_{\Theta},
\qquad
\mathcal{P}^{\,}_{\Pi}:=\openone^{\,}_{N}\otimes P^{\,}_{\Pi},
\label{eq: def reps widehat Theta and widehat Pi a}
\end{equation}
where $\openone^{\,}_{N}$ is the $N\times N$ unit matrix
and we make sure that
$P^{\,}_{\Theta}$ and $P^{\,}_{\Pi}$
represent products of transpositions so that
\begin{equation}
P^{\,}_{\Theta}= 
P^{-1}_{\Theta}=
P^{\mathsf{T}}_{\Theta},
\qquad
P^{\,}_{\Pi}=
P^{-1}_{\Pi}=
P^{\mathsf{T}}_{\Pi}.
\label{eq: def reps widehat Theta and widehat Pi a bis}
\end{equation}
Second, we introduce two column vectors 
$I^{\,}_{\Theta}\in\mathbb{Z}^{M}$ 
and 
$I^{\,}_{\Pi}\in\mathbb{Z}^{M}$,
which we combine into the two column vectors 
\begin{equation}
\mathcal{I}^{\,}_{\Theta}:=
\begin{pmatrix}
I^{\,}_{\Theta}
\\
\vdots
\\
I^{\,}_{\Theta}
\end{pmatrix},
\qquad
\mathcal{I}^{\,}_{\Pi}:=
\begin{pmatrix}
I^{\,}_{\Pi}
\\
\vdots
\\
I^{\,}_{\Pi}
\end{pmatrix},
\label{eq: def reps widehat Theta and widehat Pi b}
\end{equation}
and the $MN\times MN$ diagonal matrices
\begin{equation}
\mathcal{D}^{\,}_{\Theta}:=
\mathrm{diag}\,(\mathcal{I}^{\,}_{\Theta}),
\qquad
\mathcal{D}^{\,}_{\Pi}:=
\mathrm{diag}\,(\mathcal{I}^{\,}_{\Pi}),
\label{eq: def reps widehat Theta and widehat Pi b bis}
\end{equation}
with the components of the vectors
$\mathcal{I}^{\,}_{\Theta}$
and
$\mathcal{I}^{\,}_{\Pi}$
as diagonal matrix elements.
The vectors $I^{\,}_{\Theta}$ and $I^{\,}_{\Pi}$
are not chosen arbitrarily. We demand that
the vectors 
$(1+\mathcal{P}^{\,}_{\Theta})\,\mathcal{I}^{\,}_{\Theta}$ 
and
$(1+\mathcal{P}^{\,}_{\Pi})\,\mathcal{I}^{\,}_{\Pi}$ 
are made of even 
[for the $+1$ in Eq.~(\ref{eq: widehat Theta and widehat Phi square to pm1})]
and
odd 
[for the $-1$ in Eq.~(\ref{eq: widehat Theta and widehat Phi square to pm1})]
integer entries only, while
\begin{equation}
e^{+\mathrm{i}\pi\,\mathcal{D}^{\,}_{\Theta}}\,
\mathcal{P}^{\,}_{\Theta}=
\pm
\mathcal{P}^{\,}_{\Theta}\,
e^{+\mathrm{i}\pi\,\mathcal{D}^{\,}_{\Theta}}
\label{eq: TRS on P and I }
\end{equation}
and
\begin{equation}
e^{+\mathrm{i}\pi\,\mathcal{D}^{\,}_{\Pi}}\,
\mathcal{P}^{\,}_{\Pi}=
\pm
\mathcal{P}^{\,}_{\Pi}\,
e^{+\mathrm{i}\pi\,\mathcal{D}^{\,}_{\Pi}},
\label{eq: PHS on P and I }
\end{equation} 
in order to meet 
$\Theta^{2}=\pm1$ and $\Pi^{2}=\pm1$, respectively.
The operations of reversal of time and interchanges of
particles and holes are then represented by
\begin{align}
\widehat{\Theta}\,
\widehat{\Psi}\,
\widehat{\Theta}^{-1}
=&\,
e^{+\mathrm{i}\,\pi\,\mathcal{D}^{\,}_{\Theta}}\,
\mathcal{P}^{\,}_{\Theta}\,\widehat{\Psi},
\label{eq: def reps widehat Theta and widehat Pi c}
\\
\widehat{\Pi}\,
\widehat{\Psi}\,
\widehat{\Pi}^{-1}
=&\,
e^{+\mathrm{i}\pi\,\mathcal{D}^{\,}_{\Pi}}\,
\mathcal{P}^{\,}_{\Pi}\,
\widehat{\Psi},
\label{eq: def reps widehat Theta and widehat Pi d}
\end{align}
for the fermions and
\begin{align}
\widehat{\Theta}\,
\widehat{\Phi}\,
\widehat{\Theta}^{-1}
=&\,
\mathcal{P}^{\,}_{\Theta}\,
\widehat{\Phi}
+
\pi\,
\mathcal{K}^{-1}\,
\mathcal{I}^{\,}_{\Theta},
\label{eq: def reps widehat Theta and widehat Pi e}
\\
\widehat{\Pi}\,
\widehat{\Phi}\,
\widehat{\Pi}^{-1}
=&
\mathcal{P}^{\,}_{\Pi}\,
\widehat{\Phi}
+
\pi\,
\mathcal{K}^{-1}\,
\mathcal{I}^{\,}_{\Pi},
\label{eq: def reps widehat Theta and widehat Pi f}
\end{align}
\end{subequations}
for the bosons. One verifies that Eq.%
~(\ref{eq: widehat Theta and widehat Phi square to pm1})
is fulfilled.

Hamiltonian~(\ref{eq: hat H bosonized}) is TRS if
\begin{subequations}
\label{eq: TRS H}
\begin{equation}
\widehat{\Theta}\,
\hat{H}\,
\widehat{\Theta}^{-1}=
+\hat{H}.
\label{eq: TRS H a}
\end{equation}
This condition is met if
\begin{align}
&
P^{\,}_{\Theta}\,
V\,
P^{-1}_{\Theta}
=
+V,
\label{eq: TRS H b}
\\
&
P^{\,}_{\Theta}\,
K\,
P^{-1}_{\Theta}
=
-
K,
\label{eq: TRS H c}
\\
&
h^{\,}_{\mathcal{T}}(x)=
h^{\,}_{-\mathcal{P}^{\,}_{\Theta}\mathcal{T}}(x),
\label{eq: TRS H d}
\\
&
\alpha^{\,}_{\mathcal{T}}(x)=
\alpha^{\,}_{-\mathcal{P}^{\,}_{\Theta}\mathcal{T}}(x)
-
\pi\,
\mathcal{T}^{\mathsf{T}}\,
\mathcal{P}^{\,}_{\Theta}\,
\mathcal{I}^{\,}_{\Theta}.
\label{eq: TRS H e}
\end{align}
\end{subequations}
The 
Hamiltonian~(\ref{eq: hat H bosonized}) is PHS if
\begin{subequations}
\label{eq: PHS H}
\begin{equation}
\widehat{\Pi}\,
\hat{H}\,
\widehat{\Pi}^{-1}=
+\hat{H}.
\label{eq: PHS H a}
\end{equation}
This condition is met if (see Appendix~\ref{appendixsec: cond TRS and PHS})
\begin{align}
&
P^{\,}_{\Pi}\,
V\,
P^{-1}_{\Pi}
=
+V,
\label{eq: PHS H b}
\\
&
P^{\,}_{\Pi}\,
K\,
P^{-1}_{\Pi}
=
+
K,
\label{eq: PHS H c}
\\
&
h^{\,}_{\mathcal{T}}(x)=
h^{\,}_{+\mathcal{P}^{\,}_{\Pi}\mathcal{T}}(x),
\label{eq: PHS H d}
\\
&
\alpha^{\,}_{\mathcal{T}}(x)=
\alpha^{\,}_{\mathcal{P}^{\,}_{\Pi}\mathcal{T}}(x)
+
\pi\,
\mathcal{T}^{\mathsf{T}}\,
\mathcal{P}^{\,}_{\Pi}\,
\mathcal{I}^{\,}_{\Pi}.
\label{eq: PHS H e}
\end{align}
\end{subequations}

\subsection{Particle-hole symmetry in interacting superconductors}
\label{sec: PHS in superconductors}

The total number of fermions is a good quantum number
in any metallic or insulating phase of fermionic matter.
This is not true anymore in the mean-field treatment of superconductivity.
In a superconductor, within a mean-field approximation, charge is
conserved modulo two as Cooper pairs can be created and annihilated.
The existence of superconductors and 
the phenomenological success of the mean-field approximation
suggest that the conservation of
the total fermion number operator should be relaxed
down to its parity in a superconducting phase of matter.
If we only demand that the parity of the total fermion number is conserved,  
we may then decompose any fermionic creation operator in the position basis
into its real and imaginary parts, thereby obtaining two Hermitean operators
called Majorana operators. Any Hermitean Hamiltonian that is build out of
even powers of Majorana operators necessarily 
preserves the parity of the total fermion number operator,
but it might break the conservation of the total fermion number.
By definition, any such Hamiltonian belongs 
to the symmetry class D. 

The tool of Abelian bosonization allows to represent a fermion
operator as a single exponential of a Bose field. In Abelian bosonization,
a Majorana operator is the sum of two exponentials, and this fact makes it 
cumbersome to apply Abelian bosonization for Majorana operators.
It is possible to circumvent this difficulty by representing any
Hamiltonian from the symmetry class D in terms of the components
of Nambu spinors obeying a reality condition. 
Indeed, we may double the dimensionality of the 
single-particle Hilbert space by introducing Nambu spinors
with the understanding that (i) a reality condition on the Nambu spinors
must hold within the physical subspace of the enlarged 
single-particle Hilbert space and (ii) the dynamics dictated by the
many-body Hamiltonian must be compatible with this reality condition.%
~\cite{Altland97,CJNSP_Majorana_fields}
The reality condition keeps track of the fact that there are many 
ways to express an even polynomial of Majorana operators in terms of
the components of a Nambu spinor. The complication brought about by 
this redundency is compensated by the fact that it is straightforward 
to implement Abelian bosonization in the Nambu representation.

We implement this particle-hole doubling by assigning to every
pair of fermionic operators $\hat{\psi}$ and $\hat{\psi}^{\dag}$
(whose indices we have been omitted for simplicity)
related to each other by the reality condition
\begin{subequations}
\begin{equation}
\widehat{\Pi}\,
\hat{\psi}\,
\widehat{\Pi}^{\dag}=
\hat{\psi}^{\dag},
\end{equation}
the pair of bosonic field operators 
$\hat{\phi}$ 
and
$\hat{\phi}'$ 
related by the reality condition
\begin{equation}
\widehat{\Pi}\,
\hat{\phi}\,
\widehat{\Pi}^{\dag}=
-\hat{\phi}'.
\end{equation}
\end{subequations}
Invariance under this transformation has to be imposed on the 
(interacting) Hamiltonian in the doubled (Nambu) representation.
In addition to the PHS, we also demand, when describing the
superconducting symmetry classes, that the parity of the total fermion
number is conserved. This discrete global symmetry, 
the symmetry of the Hamiltonian under the reversal of sign of 
all fermion operators, becomes a continuous
$U(1)$ global symmetry that is responsible for the conservation 
of the electric charge in all non-superconducting symmetry classes.
In this way, all 9 symmetry classes from the tenfold way 
descend from the symmetry class D by imposing a composition of TRS, 
$U(1)$ charge conservation, and the chiral (sublattice) symmetry.

\subsection{Conditions for a spectral gap}
\label{sec: Conditions for a spectral gap}

Hamiltonian $\hat{H}^{\,}_{\mathcal{V}}$ 
in the decomposition~(\ref{eq: hat H bosonized})
has $MN$ gapless modes. 
However, $\hat{H}^{\,}_{\mathcal{V}}$  does not
commute with $\hat{H}^{\,}_{\{\mathcal{T}\}}$
and the competition between 
$\hat{H}^{\,}_{\mathcal{V}}$
and  
$\hat{H}^{\,}_{\{\mathcal{T}\}}$
can gap some, if not all, 
the gapless modes of
$\hat{H}^{\,}_{\mathcal{V}}$.
For example,
a tunneling amplitude that scatters the right mover 
into the left mover of each flavor in each wire 
will gap out the spectrum of
$\hat{H}^{\,}_{\mathcal{V}}$.

A term in $\hat{H}^{\,}_{\mathsf{\{\mathcal{T}\}}}$ has the
potential to gap out a gapless mode of $\hat{H}^{\,}_{\mathcal{V}}$ 
if 
the condition (in the Heisenberg representation)%
~\cite{Neupert11,Haldane88}
\begin{equation}
\partial^{\,}_{x}
\left[
\mathcal{T}^{\mathsf{T}}\,
\mathcal{K}\,
\widehat{\Phi}(t,x)
+
\alpha^{\,}_{\mathcal{T}}(x)
\right]=
C^{\,}_{\mathcal{T}}(x)
\label{eq: locking condition}
\end{equation}
holds for some time-independent real-valued functions $C^{\,}_{\mathcal{T}}(x)$
on the canonical momentum 
$
(4\pi)^{-1}\,\mathcal{K}\,
(\partial^{\,}_{x}\widehat{\Phi})(t,x)
$
that is conjugate to $\widehat{\Phi}(t,x)$,
when applied to the ground state. 
The
locking condition~(\ref{eq: locking condition})
removes a pair of chiral bosonic modes with opposite chiralities
from the gapless degrees of freedom of the theory.
However, not all scattering vectors $\mathcal{T}$ 
can simultaneously lead to such a locking due to quantum fluctuations. 
The set of linear combinations
$\{\mathcal{T}^{\mathsf{T}}\,\mathcal{K}\,\widehat{\Phi}(t,x)\}$
that can satisfy the locking condition%
~(\ref{eq: locking condition})
simultaneously is labeled by the subset
$\left\{\mathcal{T}\right\}^{\,}_{\mathrm{locking}}$ 
of all tunneling matrices
$\{\mathcal{T}\}$ defined by 
Eqs.~(\ref{eq: def mathcal T a})
and~(\ref{eq: def mathcal T b})
obeying the Haldane criterion~(\ref{eq: Haldane conditions})%
~\cite{Neupert11,Haldane88}
\begin{subequations}
\label{eq: Haldane conditions}
\begin{equation}
\mathcal{T}^{\mathsf{T}}\,\mathcal{K}\,\mathcal{T}=0
\end{equation}
for any 
$\mathcal{T}\in\{\mathcal{T}\}^{\,}_{\mathrm{locking}}$ 
and
\begin{equation}
\mathcal{T}^{\mathsf{T}}\,\mathcal{K}\,\mathcal{T}'=0
\end{equation}
\end{subequations}
pairwise for any
$\mathcal{T}\neq\mathcal{T}'\in\{\mathcal{T}\}^{\,}_{\mathrm{locking}}$.

\section{Reproducing the tenfold way}
\label{sec: Reproducing the tenfold way}

Our first goal is to apply the wire construction in order to reproduce
the classification of non-interacting topological insulators 
(symmetry classes A, AIII, AI, AII, BDI, and CII in Table 
\ref{table: main table})
and superconductors
(symmetry classes D, DIII, C,and CI in Table \ref{table: main table})
in ($2+1$) dimensions
(see Table~\ref{table: main table}).~\cite{Schnyder08, Kitaev09}
In this section, we will carry out the classification scheme
within the bosonized description of quantum wires. Here, we will
restrict the classification to one-body tunneling terms, i.e., $q=1$
in Eq.~(\ref{eq:q-def}), for the non-superconducting symmetry classes,
and to two-body tunneling terms, i.e., $q=2$
in Eq.~(\ref{eq:q-def}), for the superconducting symmetry classes.
In Sec.~\ref{sec: Fractionalized phases}, 
we generalize this construction to the cases 
$q>1$ and $q>2$ of multi-particle tunnelings in the non-superconducting
and superconducting symmetry classes, respectively.
The topological stability of edge modes will be an
immediate consequence of the observation that no symmetry-respecting
local terms can be added to the models that we are going to
construct.

Within the classification of non-interacting Hamiltonians,
superconductors are nothing but fermionic bilinears with a
particle-hole symmetry. The physical interpretation of the degrees of
freedom as Bogoliubov quasiparticles is of no consequence to the
analysis. In particular, they still carry an effective 
conserved $U(1)$ charge in the non-interacting description.

\subsection{Symmetry class A}
\label{subsec: Symmetry class A}

\subsubsection{SRE phases in the tenfold way}

Topological insulators in symmetry class A can be realized without any
symmetry aside from the $U(1)$ charge conservation. The wire
construction starts from wires supporting spinless fermions, so that the
minimal choice $M=2$ only counts left- and right-moving degrees of
freedom. The $K$-matrix reads
\begin{subequations}
\label{eq: def H for class A}
\begin{equation}
K:=\mathrm{diag}\,(+1,-1).
\label{eq: def H for class A a}
\end{equation}
The entry $+1$ of the $K$-matrix corresponds to a right mover.
It is depicted by the symbol $\otimes$ in
the first line of Table~\ref{table: main table}.
The entry $-1$ of the $K$-matrix corresponds to a left mover.
It is depicted by the symbol $\odot$ in
the first line of Table~\ref{table: main table}.
The operation for reversal of time in any one of the $N$ wires
is represented by 
[one verifies that Eq.~(\ref{eq: TRS on P and I }) holds]
\begin{equation}
P^{\,}_{\Theta}:=
\begin{pmatrix}
0&1
\\
1&0
\end{pmatrix},
\qquad
I^{\,}_{\Theta}:=
\begin{pmatrix}
0\\0
\end{pmatrix}.
\label{eq: def H for class A b}
\end{equation}
\end{subequations}
We define $\hat{H}^{\,}_{\{\mathcal{T}\}}$ 
by choosing $(N-1)$ scattering vectors,
whereby, for any $j=1,\cdots,(N-1)$,
\begin{subequations}
\label{eq: def H of T for class A}
\begin{equation}
\mathcal{T}^{(j)}_{(i,\gamma)}:=
\delta^{\,}_{i,j}\,
\delta^{\,}_{\gamma,2}
-
\delta^{\,}_{i-1,j}\,
\delta^{\,}_{\gamma,1}
\label{eq: def H of T for class A d}
\end{equation}
with
$i=1,\cdots,N$
and $\gamma=1,2$.
In other words,
\begin{equation}
\mathcal{T}^{(j)}:=
(0,0|\cdots|0,+1|-1,0|\cdots|0,0)^{\mathsf{T}}
\label{eq: def H of T for class A e}
\end{equation}
for $j=1,\cdots, N-1$.
Intent on helping with the interpretation of the tunneling vectors, 
we use the $|$'s in Eq.~(\ref{eq: def H of T for class A e}) to
compartmentalize the elements within a given wire. Henceforth,
there are $M=2$ vector components within each pair of $|$'s that
encode the $M=2$ degrees of freedom within a given wire. The $j$th
scattering vector~(\ref{eq: def H of T for class A e}) labels a one-body
interaction in the fermion representation that fulfills Eq.~(\ref{eq:
  def mathcal T b}) and breaks TRS, since the scattering vector
$(0,+1)^{\mathsf{T}}$ is mapped into the scattering vector
$(+1,0)^{\mathsf{T}}$ by the permutation $P^{\,}_{\Theta}$ that
represents reversal of time in a wire by exchanging right- with
left-movers.  For any $j=1,\cdots,(N-1)$, we also introduce the
amplitude
\begin{equation}
h^{\,}_{\mathcal{T}^{(j)}}(x)\geq0
\label{eq: def H of T for class A f}
\end{equation}
and the phase
\begin{equation}
\alpha^{\,}_{\mathcal{T}^{(j)}}(x)\in\mathbb{R}
\label{eq: def H of T for class A g}
\end{equation}
\end{subequations}
according to  
Eqs.~(\ref{eq: TRS H d}) and~(\ref{eq: TRS H e}), respectively.
The choices for the amplitude~(\ref{eq: def H of T for class A f})
and the phase~(\ref{eq: def H of T for class A g})
are arbitrary. In particular
the amplitude~(\ref{eq: def H of T for class A f})
can be chosen to be sufficiently large so that it is
$\hat{H}^{\,}_{\mathcal{V}}$
that may be thought of as a perturbation of 
$\hat{H}^{\,}_{\{\mathcal{T}\}}$ and not the converse.

One verifies that all $(N-1)$
scattering vectors~(\ref{eq: def H of T for class A d})
satisfy the Haldane criterion~(\ref{eq: Haldane conditions}), i.e.,
\begin{equation}
\mathcal{T}^{(i)\mathsf{T}}\,
\mathcal{K}\,
\mathcal{T}^{(j)}=0,
\qquad
i,j=1,\cdots,N-1.
\end{equation}
Correspondingly, the term $\hat{H}^{\,}_{\{\mathcal{T}\}}$
gaps out $2(N-1)$ of the $2N$ gapless modes of
$\hat{H}^{\,}_{\mathcal{V}}$. Two modes
of opposite chirality that propagate along the
first and last wire, respectively,
remain in the low energy sector of the theory. These edge
states are localized on wire $i=1$ and $i=N$, respectively,
for their overlaps with the gapped states from the bulk decay
exponentially fast as a function of the distance away from
the first and end wires. 
The energy splitting between the edge state localized on wire $i=1$
and the one localized on wire $i=N$ 
that is brought about by the bulk states vanishes
exponentially fast with increasing $N$. 
Two gapless edge states
with opposite chiralities emerge in the two-dimensional limit $N\to\infty$.

At energies much lower than the bulk gap, the effective
$\mathcal{K}$-matrix for the edge modes is
\begin{equation}
\begin{split}
\mathcal{K}^{\,}_{\mathrm{eff}}:=&\,
\mathrm{diag}(+1,0|0,0|
\cdots|0,0|0,-1).
\end{split}
\end{equation} 
Here, $\mathcal{K}^{\,}_{\mathrm{eff}}$ follows from
replacing the entries in the 
$2N\times2N$ $\mathcal{K}$ matrix
for all gapped modes by 0.
The pictorial representation of the topological phase 
in the symmetry class A with one chiral edge state per end wire 
through the wire construction 
is shown on the first row and fourth column
of Table~\ref{table: main table}.
The generalization to an arbitrary number $n$ of gapless edge states
sharing a given chirality on the first wire 
that is opposite to that of the last
wire is the following. We enlarge $M=2$ to $M=2n$ by
making $n$ identical copies of the model depicted in the first
row and fourth column of Table~\ref{table: main table}.
The stability of the $n$
chiral gapless edge states in wire $1$ and wire $N$
is guaranteed because back-scattering 
among these gapless edges state
is not allowed kinematically within wire $1$ or within wire $N$,
while back-scattering across the bulk is exponentially 
suppressed for $N$ large by locality and the gap in the bulk. 
The number of robust gapless edge states of a given chirality
is thus integer. This is the reason why $\mathbb{Z}$
is found in the third column on the first line
of Table~\ref{table: main table}.

\subsubsection{SRE phases beyond the tenfold way}

It is imperative to ask whether the phases that we constructed so far 
exhaust all possible SRE phases in the symmetry class A. 
By demanding that one-body interactions are dominant over
many-body interactions,
we have constructed all phases from the (exhaustive) classification 
for non-interacting fermions in class A and only those. 
In these phases, the same topological invariant controls the Hall and the thermal
conductivities.
However, it was observed that interacting fermion systems 
can host additional SRE phases in the symmetry class A where this connection is lost.~\cite{Lu12} 
These phases are characterized by an edge that includes charge-neutral 
chiral modes. 
While such modes contribute to the quantized energy transport 
(i.e., the thermal Hall conductivity), 
they do not contribute to the quantized charge transport 
(i.e., the charge Hall conductivity). 
By considering the thermal and 
charge Hall conductivity as two independent quantized topological responses, 
this enlarges the classification of SPT phases in the symmetry class A to 
$\mathbb{Z}\times\mathbb{Z}$.

Starting from identical fermions of charge $e$, we now construct an 
explicit wire model that stabilizes a SRE phase of matter in the symmetry class A 
carrying a non-vanishing Hall conductivity 
but a vanishing thermal Hall conductivity.
In order to build a wire-construction of such a strongly interacting 
SRE phase in the symmetry class A, 
we group three spinless electronic wires into one unit cell, i.e.,
\begin{subequations}
\begin{equation}
K:=
\mathrm{diag}(+1,-1,+1,-1,+1,-1).
\end{equation} 
It will be useful to arrange the charges $Q^{\,}_{\gamma}=1$ 
measured in units of the electron charge $e$ 
for each of the modes $\hat{\phi}^{\,}_{\gamma}$,
$\gamma=1,\cdots,M$, into a vector
\begin{equation}
Q=(1,1,1,1,1,1)^{\mathsf{T}}.
\end{equation}
\end{subequations}

The physical meaning of the tunneling vectors (interactions)
that we define below is most transparent, 
if we employ the following linear transformation on the bosonic field variables
\begin{subequations}
\label{eq: trasf for fields Phi A}
\begin{align}
&
\widehat{\Phi}(x)=: \mathcal{W}\widetilde{\Phi}(x),\\
&
\mathcal{T}=: \mathcal{W}\widetilde{\mathcal{T}},
\label{eq: T tilde if one wants to get bosons out of fermions}
\\
&
K =:  \left(W^{\mathsf{T}}\right)^{-1} \widetilde{K} W^{-1},
\\
&Q =: \left(W^{\mathsf{T}}\right)^{-1} \widetilde{Q},
\label{eq: K tilde FQHE}
\end{align}
\end{subequations}
where $\mathcal{W}$ is a $MN\times MN$ block-diagonal matrix 
with the block $W$ having integer entries and unit determinant.
The transformation $W$ that we employ is given by
\begin{equation}
W:=
\begin{pmatrix}
0&-1&-1&0&0&0\\
+1&-1&-1&0&0&0\\
+1&0&-1&0&0&0\\
0&0&0&-1&0&+1\\
0&0&0&-1&-1&+1\\
0&0&0&-1&-1&0
\end{pmatrix}.
\label{eq: trafo for bosonic chain}
\end{equation}
It brings $K$ to the form
\begin{equation}
\widetilde{K}:=
\left(
\begin{array}{cccccc}
 0 & +1 & 0 & 0 & 0 & 0 \\
 +1 & 0 & 0 & 0 & 0 & 0 \\
 0 & 0 & +1 & 0 & 0 & 0 \\
 0 & 0 & 0 & -1 & 0 & 0 \\
 0 & 0 & 0 & 0 & 0 & -1 \\
 0 & 0 & 0 & 0 & -1 & 0 \\
\end{array}
\right).
\end{equation}

As we can read off from Eq.~\eqref{eq: def hat phi's b}, the parity of
$K^{\,}_{\gamma\gamma}$ determines the self-statistics of particles of type
$\gamma=1,\cdots,N$. As Eq.~\eqref{eq: def hat phi's b}
is form invariant under the transformation
(\ref{eq: trasf for fields Phi A}),
we conclude that, with the choice~\eqref{eq: trafo for bosonic chain}, 
the transformed modes $\gamma=1,2$ as well as the modes $\gamma=5,6$ 
are pairs of bosonic degrees of freedom, 
while the third and fourth mode remain fermonic.
Furthermore, the charges transported by
the transformed modes $\widetilde{\phi}^{\,}_{\gamma}$ are given by
\begin{equation}
\widetilde{Q}=
W^{\mathsf{T}}Q=
(+2,-2,-3,-3,-2,+2)^{\mathsf{T}}.
\end{equation}

\begin{widetext} 
We may then define the charge-conserving tunneling vectors
\begin{equation}
\begin{split}
\widetilde{\mathcal{T}}^{(j)}_{1}:=&
(0,0,0,0,0,0|\cdots|0,0,+1,-1,0,0|\cdots|0,0,0,0,0,0)^{\mathsf{T}},
\qquad j=1,\cdots, N,
\\
\widetilde{\mathcal{T}}^{(j)}_{2}:=&
(0,0,0,0,0,0|\cdots|0,0,0,0,+1,0|-1,0,0,0,0,0|\cdots|0,0,0,0,0,0)^{\mathsf{T}}, 
\qquad j=1,\cdots, N-1,
\\
\widetilde{\mathcal{T}}^{(j)}_{3}:=&
(0,0,0,0,0,0|\cdots|0,0,0,0,0,+1|0,-1,0,0,0,0|\cdots|0,0,0,0,0,0)^{\mathsf{T}}, 
\qquad j=1,\cdots, N-1.
\end{split}
\label{eq: definition of interactions that produce bosons out of fermions}
\end{equation}
\end{widetext}
Using Eq.~(\ref{eq: T tilde if one wants to get bosons out of fermions}), 
these tunneling vectors can readily be rewritten in the original electronic degrees of freedom.

These tunneling vectors gap all modes in the bulk and the 
remaining gapless edge modes on the left edge are
\begin{equation}
\widetilde{K}^{\,}_{\mathrm{eff,left}}=
\begin{pmatrix}0&1\\1&0\end{pmatrix},
\qquad
\widetilde{Q}^{\,}_{\mathrm{eff,left}}=
\begin{pmatrix}
+2
\\
-2
\end{pmatrix}.
\label{eq: edge theory of interacting A}
\end{equation}
The only charge-conserving tunneling vector that could gap out this
effective edge theory, $\widetilde{T}=(1,1)^{\mathsf{T}}$, is not
compatible with Haldane's criterion~\eqref{eq: Haldane conditions}. We
conclude that the edge theory~\eqref{eq: edge theory of interacting A}
is stable against charge conserving perturbations. The Hall
conductivity supported by this edge theory is given by
\begin{equation}
\widetilde{Q}_{\mathrm{eff,left}}^{\mathsf{T}}\,
\widetilde{K}_{\mathrm{eff,left}}^{-1}\,
\widetilde{Q}_{\mathrm{eff,left}}
=-8
\end{equation}
in units of $e^2/h$. This is the minimal Hall conductivity of a SRE
phase of bosons, if each boson is interpreted as 
a pair of electrons carrying the electronic charge $2e$.~\cite{Lu12} 
On the other hand, the edge theory%
~\eqref{eq: edge theory of interacting A} 
supports two modes with opposite 
chiralities, for the symmetric matrix $\widetilde{K}^{\,}_{\mathrm{eff,left}}$
has the pair of eigenvalues $\pm1$. 
Thus, the net energy transported  along the left edge,
and with it the thermal Hall conductivity, 
vanishes.

\subsection{Symmetry class AII} 
\label{subsec: symmetry class AII}

Topological insulators in symmetry class AII can be realized by
demanding that $U(1)$ charge conservation holds and that
TRS with $\Theta^{2}=-1$ holds. The wire
construction starts from wires supporting spin-$1/2$ fermions
because $\Theta^{2}=-1$, so that the
minimal choice $M=4$ counts two pairs of Kramers degenerate
left- and right-moving degrees of freedom carrying opposite spin
projections on the spin quantization axis, i.e., 
two pairs of Kramers degenerate helical modes. 
The $K$-matrix reads
\begin{subequations}
\label{eq: def H of V for class AII}
\begin{equation}
K:=\mathrm{diag}\,(+1,-1,-1,+1).
\label{eq: def H of V for class AII a}
\end{equation}
The entries in the $K$-matrix represent, from left to right, 
a right-moving particle with spin up, 
a left-moving particle with spin down, 
a left-moving particle with spin up,  
and a right-moving particle with spin down.
The operation for reversal of time in any one of the $N$ wires
is represented by
[one verifies that Eq.~(\ref{eq: TRS on P and I }) holds]
\begin{equation}
P^{\,}_{\Theta}:=
\begin{pmatrix}
0&1&0&0
\\
1&0&0&0
\\
0&0&0&1\\
0&0&1&0
\end{pmatrix},
\qquad
I^{\,}_{\Theta}:=
\begin{pmatrix}
0\\1\\0\\1
\end{pmatrix}.
\label{eq: def H of V for class AII b}
\end{equation}
We define $\hat{H}^{\,}_{\mathcal{V}}$ 
by choosing any symmetric $4\times 4$ matrix $V$
that obeys
\begin{equation}
V=
P^{\,}_{\Theta}\,
V\,
P^{-1}_{\Theta}.
\label{eq: def H of T for class AII c}
\end{equation}
\end{subequations}
We define $\hat{H}^{\,}_{\{\mathcal{T}^{\,}_{\mathrm{SO}}\}}$ 
by choosing $2(N-1)$ scattering vectors
as follows. For any $j=1,\cdots,(N-1)$,
we introduce the pair of scattering vectors
\begin{subequations}
\begin{equation}
\mathcal{T}^{(j)}_{\mathrm{SO}}:=
(0,0,0,0|\cdots|0,0,+1,0|-1,0,0,0|\cdots|0,0,0,0)^{\mathsf{T}}
\label{eq: def H of T for class AII d}
\end{equation}
and
\begin{equation}
\overline{\mathcal{T}}^{(j)}_{\mathrm{SO}}:=
-\mathcal{P}^{\,}_{\Theta}\,
\mathcal{T}^{(j)}_{\mathrm{SO}}.
\label{eq: def H of T for class AII e}
\end{equation}
The scattering vector~(\ref{eq: def H of T for class AII d})
labels a one-body interaction in the fermion representation
that fulfills Eq.~(\ref{eq: def mathcal T b}).
It scatters a left mover with spin up from wire $j$ 
into a right mover with spin up in wire $j+1$.
For any $j=1,\cdots,(N-1)$,
we also introduce the pair of amplitudes
\begin{equation}
h^{\,}_{\mathcal{T}^{(j)}_{\mathrm{SO}}}(x)=
h^{\,}_{\overline{\mathcal{T}}^{(j)}_{\mathrm{SO}}}(x)\geq0
\label{eq: def H of T for class AII f}
\end{equation}
and the pair of phases
\begin{equation}
\alpha^{\,}_{\mathcal{T}^{(j)}_{\mathrm{SO}}}(x)=
\alpha^{\,}_{\overline{\mathcal{T}}^{(j)}_{\mathrm{SO}}}(x)\in\mathbb{R}
\label{eq: def H of T for class AII g}
\end{equation}
\end{subequations}
according to  
Eqs.~(\ref{eq: TRS H d}) and~(\ref{eq: TRS H e}), respectively.
The choices for the amplitude~(\ref{eq: def H of T for class AII f})
and the phase~(\ref{eq: def H of T for class AII g})
are arbitrary. 
The subscript SO refers to the intrinsic spin-orbit coupling.  
The rational for using it shall be shortly explained.  

One verifies that all $2(N-1)$
scattering vectors~(\ref{eq: def H of T for class AII c})
and~(\ref{eq: def H of T for class AII d})
satisfy the Haldane criterion%
~(\ref{eq: Haldane conditions}), i.e.,
\begin{equation}
\mathcal{T}^{(i)\mathsf{T}}_{\mathrm{SO}}\,
\mathcal{K}\,
\mathcal{T}^{(j)}_{\mathrm{SO}}=
\overline{\mathcal{T}}^{(i)\mathsf{T}}_{\mathrm{SO}}\,
\mathcal{K}\,
\overline{\mathcal{T}}^{(j)}_{\mathrm{SO}}=
\mathcal{T}^{(i)\mathsf{T}}_{\mathrm{SO}}\,
\mathcal{K}\,
\overline{\mathcal{T}}^{(j)}_{\mathrm{SO}}=0,
\end{equation}
for $i,j=1,\cdots,N-1$.
Correspondingly, the term $\hat{H}^{\,}_{\{\mathcal{T}^{\,}_{\mathrm{SO}}\}}$
gaps out $4(N-1)$ of the $4N$ gapless modes of
$\hat{H}^{\,}_{\mathcal{V}}$. 
Two pairs of Kramers degenerate helical edge states 
that propagate along the first and last wire,
respectively, remain in the low energy sector of the theory.
These edge states are localized on wire $i=1$ and $i=N$, respectively,
for their overlaps with the gapped states from the bulk decay
exponentially fast as a function of the distance away from
the first and end wires. 
The energy splitting between the edge state localized on wire $i=1$
and wire $i=N$ brought about by the bulk states vanishes
exponentially fast with increasing $N$. Two pairs of
gapless Kramers degenerate helical edge states
emerge in the two-dimensional limit $N\to\infty$.

At energies much lower than the bulk gap,
the effective $\mathcal{K}$-matrix for 
the two pairs of helical edge modes is
\begin{equation}
\begin{split}
\mathcal{K}^{\,}_{\mathrm{eff}}:=&\,
\mathrm{diag}(+1,-1,0,0|0,0,0,0|
\\
&\,\qquad\qquad\qquad
\cdots|0,0,0,0|0,0,-1,+1).
\end{split}
\end{equation} 
Here, $\mathcal{K}^{\,}_{\mathrm{eff}}$ follows from
replacing the entries in the 
$4N\times4N$ $\mathcal{K}$ matrix
for all gapped modes by 0.
We are going to show that
the effective scattering vector 
\begin{equation}
\mathcal{T}^{\,}_{\mathrm{eff}}:=
(+1,-1,0,0|0,0,0,0|\cdots)^{\mathsf{T}},
\label{eq: def T eff AII}
\end{equation}
with the potential to gap out the pair of 
Kramers degenerate helical edge modes on wire $i=1$
since it fulfills the Haldane criterion%
~(\ref{eq: Haldane conditions}),
is not allowed by TRS.%
~\cite{footnote:stabilityAII} 
On the one hand,
$\mathcal{T}^{\,}_{\mathrm{eff}}$
maps to itself under reversal of time,
\begin{equation}
\mathcal{T}^{\,}_{\mathrm{eff}}=
-\mathcal{P}^{\,}_{\Theta}\,
\mathcal{T}^{\,}_{\mathrm{eff}}.
\end{equation}
On the other hand,
\begin{equation}
\mathcal{T}^{\mathsf{T}}_{\mathrm{eff}}\,
\mathcal{P}^{\,}_{\Theta}\,
\mathcal{I}^{\,}_{\Theta}=
-1.
\end{equation}
Therefore, the condition~(\ref{eq: TRS H e})
for $\mathcal{T}^{\,}_{\mathrm{eff}}$ to be a TRS perturbation is not met,
for the phase $\alpha^{\,}_{\mathcal{T}^{\,}_{\mathrm{eff}}}(x)$ associated to
$\mathcal{T}^{\,}_{\mathrm{eff}}$ then obeys
\begin{equation}
\alpha^{\,}_{\mathcal{T}^{\,}_{\mathrm{eff}}}(x)=
\alpha^{\,}_{\mathcal{T}^{\,}_{\mathrm{eff}}}(x)
-
\pi,
\label{eq: TRS condition on alpha Teff class AII}
\end{equation}
a condition that cannot be satisfied.

Had we imposed a TRS with $\Theta=+1$ instead of
$\Theta=-1$ as is suited for the symmetry class AI
that describes spinless fermions with TRS, we would
only need to replace $I^{\,}_{\Theta}$ in Eq.%
~(\ref{eq: def H of V for class AII b})
by the null vector.
If so, the scattering vector~(\ref{eq: def T eff AII})
is compatible with TRS since 
the condition~(\ref{eq: TRS H e})
for TRS then becomes
\begin{equation}
\alpha^{\,}_{\mathcal{T}^{\,}_{\mathrm{eff}}}(x)=
\alpha^{\,}_{\mathcal{T}^{\,}_{\mathrm{eff}}}(x)
\end{equation}
instead of 
Eq.~(\ref{eq: TRS condition on alpha Teff class AII}).
This is the reason why symmetry class AI is 
always topologically trivial in two-dimensional space
from the point of view of the wire construction.

Note also that if we would not insist on the condition of charge neutrality 
(\ref{eq: def mathcal T b}),
the tunneling vector
 \begin{equation}
\mathcal{T}^{\prime}_{\mathrm{eff}}:=
(+1,+1,0,0|0,0,0,0|\cdots)^{\mathsf{T}},
\end{equation}
that satisfies the Haldane criterion and is compatible with TRS could gap out the 
Kramers degenerate pair of helical edge states.

\begin{widetext}
To address the question of what happens if we change $M=4$ to
$M=4n$ with $n$ any strictly positive integer
in each wire from the array, we consider, without loss of 
generality as we shall see, the case of $n=2$. To this end,
it suffices to repeat all the steps that
lead to Eq.~(\ref{eq: def T eff AII}), except for the change
\begin{equation}
\mathcal{K}^{\,}_{\mathrm{eff}}:=
\mathrm{diag}\,
(+1,-1,0,0;+1,-1,0,0|0,0,0,0;0,0,0,0|
\cdots
|0,0,0,0;0,0,0,0|0,0,-1,+1;0,0,-1,+1). 
\end{equation}
One verifies that the scattering vectors
\begin{equation}
\mathcal{T}^{\prime}_{\mathrm{eff}}:=
(+1,0,0,0;0,-1,0,0|0,0,0,0;0,0,0,0|\cdots)^{\mathsf{T}}
\label{eq: def mathcal T'eff AII}
\end{equation}
and
\begin{equation}
\mathcal{T}^{\prime\prime}_{\mathrm{eff}}:=
(0,-1,0,0;+1,0,0,0|0,0,0,0;0,0,0,0|\cdots)^{\mathsf{T}}
\label{eq: def mathcal T''eff AII}
\end{equation}
are compatible with the condition that TRS holds in that the pair
is a closed set under reversal of time,
\begin{equation}
\mathcal{T}^{\prime}_{\mathrm{eff}}=
-
\mathcal{P}^{\,}_{\Theta}\,
\mathcal{T}^{\prime\prime}_{\mathrm{eff}}.
\end{equation} 
One verifies that these scattering vectors
fulfill the Haldane criterion~(\ref{eq: Haldane conditions}).
Consequently, inclusion in $\hat{H}^{\,}_{\{\mathcal{T}^{\,}_{\mathrm{SO}}\}}$ 
of the two cosine potentials with 
$\mathcal{T}^{\prime}_{\mathrm{eff}}$ and $\mathcal{T}^{\prime\prime}_{\mathrm{eff}}$
entering in their arguments, respectively, gaps out the pair of 
Kramers degenerate helical modes on wire $i=1$. The same treatment
of the wire $i=N$ leads to the conclusion that TRS does not
protect the gapless pairs of Kramers degenerate
edge states from perturbations when $n=2$.
The generalization to 
$M=4n$ channels
is that it is only when $n$ is odd that a pair of Kramers degenerate
helical edge modes is robust to the most generic
$\hat{H}^{\,}_{\{\mathcal{T}^{\,}_{\mathrm{SO}}\}}$
of the form depicted in the fourth column on line 3 of 
Table~\ref{table: main table}. Since it is the parity of $n$
in  the number $M=4n$ of channels per wire that matters
for the stability of the Kramers degenerate helical edge states,
we use the group of two integers $\mathbb{Z}^{\,}_{2}$ under
addition modulo 2 in the third column on line 3 of 
Table~\ref{table: main table}.

\end{widetext}
If we were to impose conservation of the projection of the
spin-$1/2$ quantum number on the quantization axis, we must then
preclude from all scattering vectors processes by which a spin is flipped.
In particular, the scattering vectors%
~(\ref{eq: def mathcal T'eff AII})
and%
~(\ref{eq: def mathcal T''eff AII})
are not admissible anymore. By imposing the $U(1)$ residual symmetry
of the full $SU(2)$ symmetry group for a spin-$1/2$ degree of freedom,
we recover the group of integers $\mathbb{Z}$ under the addition
that encodes the topological stability
in the quantum spin Hall effect (QSHE).

We close the discussion of the symmetry class AII by justifying the
interpretation of the index SO as an abbreviation for the intrinsic
spin-orbit coupling. To this end, we introduce a set of 
$(N-1)$ pairs of scattering vectors
\begin{subequations}
\label{eq: def mathcal T R AII}
\begin{equation}
\mathcal{T}^{(j)}_{\mathrm{R}}:=
(0,0,0,0|\cdots|0,+1,0,0|-1,0,0,0|\cdots|0,0,0,0)^{\mathsf{T}}
\label{eq: def mathcal T R AII a}
\end{equation}
and
\begin{equation}
\overline{\mathcal{T}}^{(j)}_{\mathrm{R}}:=
-
\mathcal{P}^{\,}_{\Theta}\,
\mathcal{T}^{(j)}_{\mathrm{R}}
\label{eq: def mathcal T R AII b}
\end{equation}
for $j=1,\cdots,N-1$.
The scattering vector~(\ref{eq: def mathcal T R AII a})
labels a one-body interaction in the fermion representation
that fulfills Eq.~(\ref{eq: def mathcal T b}).
The index R is an acronym for Rashba as it describes a 
backward scattering process 
by which a left mover with spin down from wire $j$
is scattered into a right mover with spin up on wire $j+1$
and conversely.
For any $j=1,\cdots,(N-1)$,
we also introduce the pair of amplitudes
\begin{equation}
h^{\,}_{\mathcal{T}^{(j)}_{\mathrm{R}}}(x)=
h^{\,}_{\overline{\mathcal{T}}^{(j)}_{\mathrm{R}}}(x)\geq0
\label{eq: def mathcal T R AII c}
\end{equation}
and the pair of phases
\begin{equation}
\alpha^{\,}_{\mathcal{T}^{(j)}_{\mathrm{R}}}(x)=
\alpha^{\,}_{\overline{\mathcal{T}}^{(j)}_{\mathrm{R}}}(x)
+
\pi
\in\mathbb{R}
\label{eq: def mathcal T R AII d}
\end{equation}
\end{subequations}
according to  
Eqs.~(\ref{eq: TRS H d}) and~(\ref{eq: TRS H e}), respectively.
In contrast to the intrinsic spin-orbit scattering vectors,
the Rashba scattering vectors~(\ref{eq: def mathcal T R AII a})
fail to meet the Haldane criterion~(\ref{eq: Haldane conditions})
as
\begin{equation}
\mathcal{T}^{(j)\mathsf{T}}_{\mathrm{R}}\,
\mathcal{K}\,
\overline{\mathcal{T}}^{(j+1)}_{\mathrm{R}}=
-1,
\qquad
j=1,\cdots, N-1.
\end{equation}
Hence, the Rashba scattering processes fail to open a gap in the bulk,
as is expected of a Rashba coupling in a two-dimensional electron gas.
On the other hand, the intrinsic spin-orbit coupling can
lead to a 
phase with a gap in the bulk that supports the spin quantum Hall effect
in a two-dimensional electron gas.

\subsection{Symmetry class D}
\label{subsec: Symmetry class D}

The simplest example among the topological superconductors 
can be found in the symmetry class D that is defined by the presence
of a PHS with $\Pi^{2}=+1$ and the absence of TRS.

With the understanding of PHS as discussed in Sec.~\ref{sec: PHS in superconductors}, 
we construct a representative phase in class D from identical wires supporting
right- and left-moving
spinless fermions each of which carry a particle or a hole label,
i.e., $M=4$. The $K$-matrix reads
\begin{subequations}
\label{eq: def H of V for class D}
\begin{equation}
K:=
\mathrm{diag}(+1,-1,-1,+1).
\label{eq: def H of V for class D a}
\end{equation}
The entries in the $K$-matrix represent, from left to right, 
a right-moving particle, 
a left-moving particle, 
a left-moving hole, 
and a right-moving hole.
The operation for the exchange of particles and holes 
in any one of the $N$ wires is represented by 
[one verifies that Eq.~(\ref{eq: PHS on P and I }) holds]
\begin{equation}
P^{\,}_{\Pi}:=
\begin{pmatrix}
0&0&0&1
\\
0&0&1&0
\\
0&1&0&0
\\
1&0&0&0
\end{pmatrix},
\qquad
I^{\,}_{\Pi}:=
\begin{pmatrix}
0
\\
0
\\
0
\\
0
\end{pmatrix}.
\label{eq: def H of V for class D b}
\end{equation}
We define $\hat{H}^{\,}_{\mathcal{V}}$ 
by choosing any symmetric $4\times4$ matrix $V$ that obeys
\begin{equation}
V=
+
P^{\,}_{\Pi}\,
V\,
P^{-1}_{\Pi}.
\label{eq: def H of V for class D c}
\end{equation}
\end{subequations}
We define $\hat{H}^{\,}_{\{\mathcal{T}\}}$ 
by choosing $2N-1$ scattering vectors
as follows. For any wire $j=1,\cdots,N$,
we introduce the scattering vector
\begin{subequations}
\label{eq: def H of T for class D}
\begin{equation}
\mathcal{T}^{(j)}:=
(0,0,0,0|\cdots|+1,-1,-1,+1|\cdots|0,0,0,0)^{\mathsf{T}}.
\label{eq: def H of T for class D d}
\end{equation}
Between any pair of neighboring wires 
we introduce the scattering vector
\begin{equation}
\begin{split}
&
\overline{\mathcal{T}}^{(j)}:=
\\
&
(0,0,0,0|\cdots|0,+1,-1,0|-1,0,0,+1|\cdots|0,0,0,0)^{\mathsf{T}},
\end{split}
\label{eq: def H of T for class D e}
\end{equation}
for $j=1,\cdots,(N-1)$.
We observe that both $\mathcal{T}^{(j)}$ and $\overline{\mathcal{T}}^{(j)}$
are eigenvectors of the particle-hole transformation in that
\begin{equation}
\mathcal{P}^{\,}_{\Pi}\,
\mathcal{T}^{(j)}=
+\mathcal{T}^{(j)},
\qquad
\mathcal{P}^{\,}_{\Pi}\,
 \overline{\mathcal{T}}^{(j)}=
-\overline{\mathcal{T}}^{(j)}.
\end{equation}
Thus, to comply with PHS, we have to demand that the phases
\begin{equation}
\alpha^{\,}_{\overline{\mathcal{T}}^{(j)}}(x)
=0,
\label{eq: def H of T for class D g}
\end{equation}
\end{subequations}
while
$\alpha^{\,}_{\mathcal{T}^{(j)}}(x)$ are unrestricted.
Similarly, the amplitudes $h^{\,}_{\mathcal{T}^{(j)}}(x)$
and $h^{\,}_{\overline{\mathcal{T}}^{(j)}}(x)$ can take arbitrary real values.

One verifies that the set of scattering vectors defined by
Eqs.~(\ref{eq: def H of T for class D d}) and~
(\ref{eq: def H of T for class D e}) 
satisfies the Haldane criterion.  
Correspondingly, the term
$\hat{H}^{\,}_{\{\mathcal{T}\}}$ gaps out $(4N-2)$ of the $4N$ gapless
modes of $\hat{H}^{\,}_{\mathcal{V}}$. Furthermore, one identifies with
\begin{equation}
\overline{\mathcal{T}}^{(0)}
=(-1,0,0,+1|0,0,0,0|\cdots|0,0,0,0|0,+1,-1,0)^{\mathsf{T}}
\end{equation} 
a unique (up to an integer multiplicative factor)
scattering vector that satisfies the Haldane criterion with all
existing scattering vectors Eqs.~(\ref{eq: def H of T for class D d})
and~(\ref{eq: def H of T for class D e}) and could thus potentially
gap out the remaining pair of modes. However,
the tunneling $\overline{\mathcal{T}}^{(0)}$ is non-local 
for it connects the two edges of the system
when open boundary conditions are chosen.
We thus conclude that the two remaining modes are exponentially
localized near wire $i=1$ and wire $i=N$, respectively, and propagate
with opposite chirality.

To give a physical interpretation of the resulting topological (edge)
theory in this wire construction, one has to keep in mind that the
degrees of freedom were artificially doubled.  We found, in this
doubled theory, a single chiral boson (with chiral central charge
$c=1$). To interpret it as the edge of a chiral 
$(p^{\,}_{x}+\mathrm{i}p^{\,}_{y})$
superconductor, we impose the reality condition to obtain a single
chiral Majorana mode with chiral central charge $c=1/2$.

The pictorial representation of the topological phase 
in the symmetry class D through the wire construction 
is shown on the fifth row of Table~\ref{table: main table}. 
The generalization to an arbitrary number $n$ of 
gapless chiral edge modes
is analogous to the case discussed in symmetry class A. 
The number of robust gapless chiral edge states of a given chirality is thus integer. 
This is the reason why the group of integers
$\mathbb{Z}$ is found in the third column on the fifth line
of Table~\ref{table: main table}.

\subsection{Symmetry classes DIII and C}

The remaining two topological nontrivial superconducting classes DIII
(TRS with $\Theta^{2}=-1$ and PHS with $\Pi^{2}=+1$) 
and C (PHS with $\Pi^{2}=-1$)
involve spin-$1/2$ fermions. Each wire thus features no less than
$M=8$ internal degrees of freedom corresponding to the spin-$1/2$,
chirality, and particle/hole indices. The construction is very similar
to the cases we already presented. We relegate details to the appendices
\ref{appendixsec: Symmetry class C} and 
\ref{appendixsec: Symmetry class DIII}.

The scattering vectors that are needed to gap out the bulk for each class of class DIII and C are represented pictorially in the
fourth column on lines 4 and 9
of Table~\ref{table: main table}.

\subsection{Summary}

We have provided an explicit construction by way of an array of wires
supporting fermions that realizes all five insulating and
superconducting topological phases of matter with a nondegenerate
ground state in two-dimensional space according to the tenfold
classification of band insulators and superconductors. The topological
protection of edge modes in the bosonic formulation follows from
imposing the Haldane criterion~(\ref{eq: Haldane conditions}) along
with the appropriate symmetry constraints. 
In the next section we shall
extend the wire construction to allow many-body tunneling processes
that delivers fractionalized phases with degenerate ground states.

\section{Fractionalized phases}
\label{sec: Fractionalized phases}

The power of the wire construction goes much beyond what we have used 
in Sec.~\ref{sec: Reproducing the tenfold way} 
to reproduce the classification of the SRE phases.
In this section we describe how to construct models for interacting
phases of matter with intrinsic topological order and fractionalized
excitations by relaxing the condition on the tunnelings 
between wires that they be of the one-body type.
While these phases are more complex, the principles for
constructing the models and proving the stability of edge modes
remain the same: All allowed tunneling vectors have to obey the Haldane
criterion~(\ref{eq: Haldane conditions}) and the respective
symmetries. 

\subsection{Symmetry class A: Fractional quantum Hall states}
\label{subsec: Symmetry class A: Fractional quantum Hall states}

First, we review the models of quantum wires
that are topologically equivalent to the
Laughlin state in the FQHE,%
~\cite{Laughlin83}
following the construction in Ref.~\onlinecite{Kane02} 
for Abelian fractional quantum Hall states.
Here, we want to
emphasize that the choice of scattering vectors is determined by
the Haldane criterion~(\ref{eq: Haldane conditions})
and at the same time prepare the grounds for the
construction of fractional topological insulators with TRS in
Sec.~\ref{subsec: Symmetry Class AII: Fractional topological insulators}.

We want to construct the fermionic Laughlin series of states indexed by 
the positive odd integer $m$.~\cite{Laughlin83}
(By the same method, other fractional quantum Hall phases
from the Abelian hierarchy could be constructed.%
\cite{Kane02})
The elementary degrees of freedom in each wire are spinless 
right- and left-moving fermions with the $K$-matrix
\begin{subequations}
\label{eq: def array quantum wires TSE A} 
\begin{equation}
K=
\mathrm{diag}\,(+1,-1),
\label{eq: def array quantum wires TSE A a} 
\end{equation}
as is done in 
Eq.~(\ref{eq: def H for class A a}). 
Reversal of time is defined through
$P^{\,}_{\Theta}$ and $I^{\,}_{\Theta}$ given in
Eq~(\ref{eq: def H for class A b}). Instead of
Eq~(\ref{eq: def H of T for class A}), 
the scattering vectors that describe the interactions
between the wires are now defined by
\begin{equation}
\mathcal{T}^{(j)}:=
\left(
0, 0
\left|
\cdots
\left|
m^{\,}_{+},
-m^{\,}_{-}
\left|
m^{\,}_{-},
-m^{\,}_{+}
\right.
\right|
\cdots
\right|
0,0
\right)^{\mathsf{T}},
\label{eq: def array quantum wires TSE A b} 
\end{equation}
\end{subequations}
for any $j=1,\cdots,N-1$,
where $m^{\,}_{\pm}=(m\pm 1)/2$
[see Table~\ref{table: main table} for an
illustration of the scattering process]. 

For any $j=1,\cdots,N-1$,
the scattering (tunneling) vectors%
~(\ref{eq: def array quantum wires TSE A b})
preserve the conservation of the total fermion number 
in that they obey Eq.~\eqref{eq: def mathcal T b},
and they encode a tunneling interaction of order $q=m$, with $q$ defined in Eq.~\eqref{eq:q-def}.
As a set, all tunneling interactions satisfy
the Haldane criterion~(\ref{eq: Haldane conditions}), for
\begin{equation}
\begin{split}
\mathcal{T}^{(i)\mathsf{T}}\,
\mathcal{K}\,
\mathcal{T}^{(j)}
=&\,
0,
\quad
i,j=1,\cdots,N-1.
\end{split}
\label{eq: def Tj for A}
\end{equation} 
We note that the choice of tunneling vector in 
Eq.~\eqref{eq: def array quantum wires TSE A b} 
is unique (up to an integer multiplicative factor)
if one insists on charge conservation,
compliance with the Haldane
criterion~(\ref{eq: Haldane conditions}), and only includes scattering
between neighboring wires. 

The bare counting of tunneling vectors shows that the wire model gaps
out all but two modes. However, we still have to convince ourself that
the remaining two modes (i) live on the edge, (ii) cannot be gapped
out by other (local) scattering vectors and (iii) are made out of
fractionalized quasiparticles.

To address (i) and (ii), we note that the remaining two modes can be
gapped out by a unique (up to an integer multiplicative factor)
charge-conserving scattering vector that
satisfies the Haldane criterion~(\ref{eq: Haldane conditions}) 
with all existing scatterings, namely
\begin{equation}
\mathcal{T}^{(0)}:=
\left(\left.\left.
m^{\,}_{-},
-m^{\,}_{+}
\right|
0,
0
\right|
\cdots
\left|
0,
0
\left|
m^{\,}_{+},
-m^{\,}_{-}
\right.\right.
\right)^{\mathsf{T}}.
\label{eq: def T0 for A}
\end{equation}
Connecting the opposite ends of the array of wires 
through the tunneling
$\mathcal{T}^{(0)}$ is not an admissible perturbation,
for it violates locality in the 
two-dimensional thermodynamic limit $N\to\infty$.
Had we chosen periodic boundary conditions 
corresponding to a cylinder geometry (i.e., a tube as in
Fig.~\ref{fig: BoundaryConditions}) by which 
the first and last wire are nearest neighbors,
$\mathcal{T}^{(0)}$ would be admissible. 
Hence, the gapless nature of the remaining modes 
when open boundary conditions are chosen depends
on the boundary conditions. These gapless modes have support 
near the boundary only and are topologically protected.

Applying the transformation~\eqref{eq: trasf for fields Phi A} with
\begin{equation}
W:=
\begin{pmatrix}
-m^{\,}_{-}&m^{\,}_{+}
\\
m^{\,}_{+}&-m^{\,}_{-}
\end{pmatrix},
\label{eq: trasf for fields Phi A b R}
\end{equation}
where 
\begin{equation}
\mathrm{det}\,W=-m,
\end{equation}
transforms the $K$-matrix into
\begin{equation}
\begin{split}
\widetilde{K}
=&\,
\begin{pmatrix}
-m&0
\\
0&+m
\end{pmatrix}.
\end{split}
\end{equation}
As its determinant is not unity, 
the linear transformation~(\ref{eq: trasf for fields Phi A b R})
changes the compactification radius of the new field $\widetilde{\Phi}(x)$ 
relative to the compactification radius of the old field $\widehat{\Phi}(x)$ 
accordingly. Finally, the transformed tunneling vectors are given by
\begin{align}
\widetilde{\mathcal{T}}^{(j)}
=&\,
(0,0|\cdots|0,0|0,+1|-1,0|0,0|\cdots|0,0)^{\mathsf{T}},
\label{eq: T tilde FQHE}
\end{align}
where $\mathcal{W}:=\openone^{\,}_{N}\otimes W$ and
$j=1,\cdots,N-1$.

In view of Eqs.~\eqref{eq: K tilde FQHE}
and~\eqref{eq: T tilde FQHE}, the remaining effective edge theory is
described by
\begin{equation}
\widetilde{\mathcal{K}}_{\mathrm{eff}}=
\mathrm{diag}\,
(-m,0|0,0|\cdots|0,0|0,+m).
\label{eq: effective K FQHE}
\end{equation} 
This is a chiral theory at each edge that cannot be gapped by local
perturbations. Equation~\eqref{eq: effective K FQHE} is precisely the
edge theory for anyons with statistical angle $1/m$ and charge $e/m$,%
~\cite{Wen91}
where $e$ is the charge of the original fermions.

\subsection{Symmetry Class AII: Fractional topological insulators}
\label{subsec: Symmetry Class AII: Fractional topological insulators}

Having understood how fractionalized quasiparticles emerge out of a
wire construction, it is imperative to ask what other phases can be
obtained when symmetries are imposed on the topologically ordered
phase.  Such symmetry enriched topological phases have been classified
by methods of group cohomology.%
~\cite{Chen13}
Here, we shall exemplify
for the case of TRS with $\Theta^{2}=-1$ 
how the wire construction can be used to build up an intuition 
for these phases and to study the stability of their edge theory.

The elementary degrees of freedom in each wire are spin-$1/2$ 
right- and left-moving fermions with the $K$-matrix
\begin{subequations} 
\label{eq: def array quantum wires TSE AII} 
\begin{equation}  
K:=
\mathrm{diag}\,(+1,-1,-1,+1),
\label{eq: def array quantum wires TSE AII a}  
\end{equation}
as is done in Eq.~(\ref{eq: def H of V for class AII a}).
Reversal of time is defined through
$P^{\,}_{\Theta}$ and $I^{\,}_{\Theta}$ given in
Eq~(\ref{eq: def H of V for class AII b}). Instead of
Eq~(\ref{eq: def H of T for class AII d}), 
the scattering vectors that describe the interactions between
the wires are now defined by
\begin{widetext}
\begin{equation}
\mathcal{T}^{(j)}:=
\left(
0,0,0,0
\left|
\cdots
\left|
-m^{\,}_{-},
0,
+m^{\,}_{+},
0
\left|
-m^{\,}_{+},
\right.
0,
+m^{\,}_{-},
0
\right|
\cdots
\right|
0,0,0,0
\right)^{\mathsf{T}}
\label{eq: def array quantum wires TSE AII b} 
\end{equation}
and
\begin{equation}
\overline{\mathcal{T}}^{(j)}:=
-
\mathcal{P}^{\,}_{\Theta}\,\mathcal{T}^{(j)},
\label{eq: def array quantum wires TSE AII c} 
\end{equation}
\end{widetext}
\end{subequations}
for any $j=1,\cdots,N-1$, $m$ a positive odd integer, and $m^{\,}_{\pm}=(m\pm1)/2$.

For any $j=1,\cdots,N-1$,
the scattering (tunneling) vectors%
~(\ref{eq: def array quantum wires TSE AII b} )
preserve conservation of the total fermion number in that
they obey Eq.~\eqref{eq: def mathcal T b},
and they encode a tunneling interaction of order $q=m$ with $q$ defined in Eq.~\eqref{eq:q-def}.
They also satisfy the Haldane criterion~(\ref{eq: Haldane conditions})
as a set  
[see Table~\ref{table: main table} for an
illustration of the scattering process].

Applying the transformation~\eqref{eq: trasf for fields Phi A} with 
\begin{equation}
W:=
\begin{pmatrix}
-m^{\,}_{-}&0&m^{\,}_{+}&0\\
0&-m^{\,}_{-}&0&m^{\,}_{+}\\
m^{\,}_{+}&0&-m^{\,}_{-}&0\\
0&m^{\,}_{+}&0&-m^{\,}_{-}
\end{pmatrix},
\label{eq: linear trafo for FTI}
\end{equation}
to the bosonic fields,
leaves the representation of time-reversal invariant 
\begin{equation}
W^{-1}\,P^{\,}_{\Theta}\,W=P^{\,}_{\Theta},
\label{eq: W commutes with P SEP class AII}
\end{equation}
while casting the theory in a new form with the transformed
$\widetilde{K}$-matrix given by
\begin{equation}
\widetilde{K}=
\mathrm{diag}\,(-m,+m,+m,-m),
\end{equation}
and, for any $j=1,\cdots,N-1$,
with the transformed pair of scattering vectors 
$(\widetilde{\mathcal{T}}^{j},\widetilde{\overline{\mathcal{T}}}^{j})$
given by
\begin{widetext}
\begin{equation}
\widetilde{\mathcal{T}}^{(j)}=
(
0,0,0,0
|
\cdots
|
+1,0,0,0
|
0,0,-1,0
|
\cdots
|
0,0,0,0
)^{\mathsf{T}}
\end{equation}
and
\begin{equation}
\widetilde{\overline{\mathcal{T}}}^{(j)}=
(
0,0,0,0
|
\cdots
|
0,-1,0,0
|
0,0,0,+1
|
\cdots
|
0,0,0,0
)^{\mathsf{T}}.
\end{equation}
When these scattering vectors have gapped out all modes in the bulk,
the effective edge theory is described by
\begin{equation}
\widetilde{\mathcal{K}}_{\mathrm{eff}}=
\mathrm{diag}\,
(0,0,+m,-m|
0,0,0,0|
\cdots
|0,0,0,0|
-m,+m,0,0).
\label{eq: effective K}
\end{equation}
\end{widetext}
This effective $K$-matrix describes 
a single Kramers degenerate pair of $1/m$ anyons propagating along
the first wire and another 
single Kramers degenerate pair of $1/m$ anyons propagating along
the last wire. Their robustness to local perturbations
is guaranteed by TRS.

Unlike in the tenfold way, the correspondence between the bulk
topological phase and the edge theories of LRE phases is not
one-to-one. For example, while a bulk topological LRE phase supports
fractionalized topological excitations in the bulk, its edge modes
may be gapped out by symmetry-allowed perturbations. For the phases
discussed in this section, namely the Abelian and TRS fractional
topological insulators, it was shown in Refs.~\onlinecite{Neupert11}
and \onlinecite{Levin09} that the edge, consisting of Kramers
degenerate pairs of edge modes, supports at most one stable Kramers
degenerate pair of delocalized quasiparticles that are stable
against disorder.  (Note that this does not preclude the richer edge
physics of non-Abelian TRS fractional topological
insulators.~\cite{scharfenberger})

We will now argue that the wire constructions with edge modes
given by Eq.~\eqref{eq: effective K} exhaust all stable edge
theories of Abelian topological phases which are protected by TRS
with $\Theta^{2}=-1$ alone.

Let the single protected Kramers degenerate pair be characterized 
by the linear combination of bosonic fields
\begin{equation}
\hat{\varphi}(x):=
\mathcal{T}^{\mathsf{T}}\,
\mathcal{K}'\,
\widehat{\Phi}(x)
\end{equation}
and its time-reversed partner
\begin{equation}
\hat{\bar{\varphi}}(x):=
\overline{\mathcal{T}}^{\mathsf{T}}\,
\mathcal{K}'\,
\widehat{\Phi}(x),
\end{equation}
where the tunneling vector $\mathcal{T}$ was constructed from the
microscopic information from the theory in Ref.~\onlinecite{Neupert11}
and $\mathcal{K}'$ is the $K$-matrix of a TRS bulk Chern-Simons
theory from the theory in Ref.~\onlinecite{Neupert11}. 
[In other words, the theory encoded by $\mathcal{K}'$
has nothing to do a priori with the array of quantum wires defined
by Eq.~(\ref{eq: def array quantum wires TSE AII}).]  
The Kramers degenerate pair of modes 
$(\hat{\varphi},\hat{\bar{\varphi}})$ 
is stable against TRS perturbations supported on a single edge 
if and only if
\begin{equation}
\frac12|\mathcal{T}^{\mathsf{T}}\,\mathcal{Q}|
\end{equation}
is an odd number. Here, $\mathcal{Q}$ is the charge vector with integer entries that
determines the coupling of the different modes to the electromagnetic
field. Provided $(\hat{\varphi},\hat{\bar{\varphi}})$ 
is stable, its equal-time
commutation relations follow from 
Eq.~(\ref{eq: def hat phi's b})
as  
\begin{subequations}
\begin{align}
\left[\hat{\varphi}(x),\hat{\varphi}(x')\right]
=&\,
-\mathrm{i}\pi\,
\left(
\mathcal{T}^{\mathsf{T}}\,
\mathcal{K}'\,
\mathcal{T}\,
\mathrm{sgn}(x-x')
+
\mathcal{T}^{\mathsf{T}}\,
\mathcal{L}\,
\mathcal{T}
\right),
\\
\left[\hat{\bar{\varphi}}(x),\hat{\bar{\varphi}}(x')\right]
=&\,
-\mathrm{i}\pi\,
\left(
-\mathcal{T}^{\mathsf{T}}\,
\mathcal{K}'\,
\mathcal{T}\,
\mathrm{sgn}(x-x')
+
\overline{\mathcal{T}}^{\mathsf{T}}\,
\mathcal{L}\,
\overline{\mathcal{T}}
\right),
\end{align}
\end{subequations} 
where we used that $\mathcal{K}'$ anticommutes with 
$\mathcal{P}^{\,}_{\Theta}$ 
according to Eq.~(\ref{eq: TRS H c}). 
By the same token, 
we can show that the fields 
$\hat{\varphi}$ and 
$\hat{\bar{\varphi}}$ commute, 
for
\begin{equation}
\mathcal{T}^{\mathsf{T}}\,
\mathcal{K}'\,
\overline{\mathcal{T}}=
\mathcal{T}^{\mathsf{T}}\,
\mathcal{P}^{\,}_{\Theta}\,
\mathcal{K}'\,\mathcal{T}=
-
\overline{\mathcal{T}}^{\mathsf{T}}\,
\mathcal{K}'\,
\mathcal{T}=0.
\end{equation}
We conclude that the effective edge theory for \emph{any} Abelian TRS
fractional topological insulator build from fermions has the effective
form of one Kramers degenerate pairs
\begin{equation}
\mathcal{K}_{\mathrm{eff}}=
\begin{pmatrix}
\mathcal{T}^{\mathsf{T}}\mathcal{K}'\mathcal{T}&0\\
0&-\mathcal{T}^{\mathsf{T}}\mathcal{K}'\mathcal{T}
\end{pmatrix},
\end{equation}
and is thus entirely defined by the single integer
\begin{equation}
m:=\mathcal{T}^{\mathsf{T}}\mathcal{K}'\mathcal{T}.
\end{equation}
With the scattering vectors%
~(\ref{eq: def array quantum wires TSE AII c}) 
we have given an explicit wire construction for each of these
cases, thus exhausting all possible stable edge theories for Abelian
fractional topological insulators.

For each positive odd integer $m$, we can thus say that 
the fractionalized mode has a $\mathbb{Z}^{\,}_{2}$ character: 
It can have either one or none stable Kramers degenerate pair of $m$ 
quasiparticles. 

\subsection{Symmetry Class D: Fractional superconductors}
\label{subsec: Symmetry Class D: Fractional superconductors}

In Sec.%
~\ref{subsec: Symmetry Class AII: Fractional topological insulators} 
we have imposed TRS on the wire
construction of fractional quantum Hall states and obtained the
fractional topological insulator in symmetry class AII.  In complete
analogy, we can impose PHS with
$\Pi^{2}=+1$ on the wire construction of a fractional quantum
Hall state, thereby promoting it to symmetry class D.  Physically,
there follows a model for a superconductor with ``fractionalized" Majorana fermions or
Bogoliubov quasiparticles. 

Lately, interest in this direction has been revived by the investigation 
of exotic quantum dimensions of twist defects embedded 
in an Abelian fractional quantum Hall liquid,%
~\cite{maissam1,maissam2,maissam3} 
along with heterostructures of superconductors combined 
with fractional quantum Hall effect,%
~\cite{abolhassan,netanel,jason1} 
or fractional topological insulators.%
~\cite{cheng1} 
Furthermore, the Kitaev quantum wire has been generalized to 
$\mathbb{Z}^{\,}_{n}$ clock models hosting parafermionic edge modes,%
~\cite{kitaev,fendley}
along with efforts to transcend the
Read-Rezayi quantum Hall state%
~\cite{nicanded} 
to spin liquids%
~\cite{nacsl1,nacsl2} 
and superconductors,%
~\cite{gangof11} 
all of which exhibit parafermionic quasiparticles.

As in the classification of non-interacting
insulators, we treat the Bogoliubov quasiparticles under bosonization
as if they were Dirac fermions. The  fractional 
phase is driven by interactions among the Bogoliubov quasiparticles. 

The elementary degrees of freedom in each wire are spinless right- and
left-moving fermions and holes as was defined for symmetry class D in
Eqs.~\eqref{eq: def H of V for class D a}%
-\eqref{eq: def H of V for class D c}.  
We construct the fractional
topological insulator using the set of PHS scattering vectors 
$\mathcal{T}^{(j)}$ , for $j=1,\cdots,N$
with $\mathcal{T}^{(j)}$ as defined in 
Eq.~\eqref{eq: def H of T for class D d} in each wire
and the PHS as defined in Eq.~\eqref{eq: def H of V for class D b}.
We complement them with the set of PHS scattering vectors 
$\overline{\mathcal{T}}^{(j)}$, for $j=1,\cdots,N-1$ defined by
\begin{widetext}
\begin{equation}
\overline{\mathcal{T}}^{(j)}
=\left(
0,0,0,0\left|
\cdots
\left|\left.
-m^{\,}_{-},m^{\,}_{+},-m^{\,}_{+}, m^{\,}_{-}
\right|
-m^{\,}_{+},m^{\,}_{-},-m^{\,}_{-}, m^{\,}_{+}
\right|
\cdots
\right|0,0,0,0
\right)^{\mathsf{T}},
\qquad
m^{\,}_{\pm}=(m\pm1)/2,
\label{eq: def overline T for class D j=1 ... N-1}
\end{equation}
with $m$ an odd positive integer.  Notice
that 
$\overline{\mathcal{T}}^{(j)}:=
-\mathcal{P}_{\Pi}\,\overline{\mathcal{T}}^{(j)}$ so that we have to
demand that $\alpha^{\,}_{\overline{\mathcal{T}}^{(j)}}=0$ has to comply with PHS. 
Thus, together the $\mathcal{T}^{(j)}$ and $\overline{\mathcal{T}}^{(j)}$
gap out $(4N-2)$ of the $4N$ chiral modes in the wire. We can
identify a unique (up to an integer multiplicative factor) 
scattering vector
\begin{equation}
\overline{\mathcal{T}}^{(0)}=
\left(\left.\left.
-m^{\,}_{+},m^{\,}_{-},-m^{\,}_{-}, m^{\,}_{+}
\right|0,0,0,0\right|
\cdots
\left|0,0,0,0\left|
-m^{\,}_{-},m^{\,}_{+},-m^{\,}_{+}, m^{\,}_{-}
\right.\right.
\right)^{\mathsf{T}},
\qquad
m^{\,}_{\pm}=(m\pm1)/2,
\label{eq: T0 for SET D}
\end{equation}
with $m$ the ame odd positive integer as in Eq.\ (\ref{eq: def
  overline T for class D j=1 ... N-1}) that satisfies the Haldane
criterion with all $\mathcal{T}^{(j)}$ and
$\overline{\mathcal{T}}^{(j)}$ and thus can potentially gap out the 2
remaining modes. However, it is physically forbidden for it represents
a non-local scattering from one edge to the other. We conclude that
each boundary supports a single remaining chiral mode that is an
eigenstate of PHS.
\end{widetext}

To understand the nature of the single remaining chiral mode on each boundary, we use the  local linear transformation $W$ of the bosonic fields
\begin{equation}
W=
\begin{pmatrix}
-m^{\,}_{-}&+m^{\,}_{+}&0&0\\
+m^{\,}_{+}&-m^{\,}_{-}&0&0\\
0&0&-m^{\,}_{-}&+m^{\,}_{+}\\
0&0&+m^{\,}_{+}&-m^{\,}_{-}
\end{pmatrix},
\quad
m^{\,}_{\pm}=\frac{m\pm1}{2},
\end{equation}
with determinant $\mathrm{det}\,W=m^{4}$.
When applied to the non-local scattering vector $\overline{\mathcal{T}}^{(0)}
$ that connects the two remaining chiral edge modes, 
\begin{equation}
\begin{split}
\widetilde{\overline{\mathcal{T}}}^{(0)}&=\mathcal{W}^{-1}\,\overline{\mathcal{T}}^{(0)}
\\&
=(0,-1,+1,0|0,0,0,0|\cdots|0,0,0,0|+1,0,0,-1),
\end{split}
\end{equation}
while the $K$ matrix changes under this transformation to
\begin{equation}
\widetilde{K}=
\mathrm{diag}\,(-m,m,m,-m).
\end{equation}
Noting that the representation of PHS is unchanged \begin{equation}
W^{-1}\,P^{\,}_{\Pi}\,W=P^{\,}_{\Pi},
\label{eq: W commutes with P SEP class D}
\end{equation}
we can interpret the remaining chiral edge mode as a PHS superposition
of a Laughlin quasiparticle and a Laughlin quasihole.  It thus 
describes a fractional chiral
edge mode on either side of the
two-dimensional array of quantum wires.
The definite chirality is
an important difference to the case of the fractional
$\mathbb{Z}^{\,}_{2}$ topological insulator discussed in
Sec.~\ref{subsec: Symmetry Class AII: Fractional topological
  insulators}.  It guarantees that any integer number $n\in\mathbb{Z}$
layers of this theory is stable, for no tunneling vector that acts
locally on one edge can satisfy the Haldane criterion~(\ref{eq:
  Haldane conditions}). For each $m$, we can thus say that the
parafermion mode has a $\mathbb{Z}$ character, as does the SRE phase
in symmetry class D.

\subsection{Symmetry classes DIII and C:
More fractional superconductors  }

Needed are the many-body tunneling matrices for class
DIII and C. We refer the reader to the appendices
\ref{appendixsec: Symmetry class C} and 
\ref{appendixsec: Symmetry class DIII}
for their definitions.
For class DIII, the edge excitations (and bulk quasiparticles) 
of the phase are TRS fractionalized Bogoliubov quasiparticles
that have also been discussed in one-dimensional realizations.
(In the latter context, these TRS 
fractionalized Bogoliubov quasiparticles are
rather susceptible to perturbations.~\cite{sela,Klinovaja13})

\section{Discussion}
\label{sec: Discussion}

In this work, we have developed a wire construction to build models of 
short-range entangled and long-range entangled topological phases in 
two spatial dimensions, so as to yield immediate information about 
the topological stability of their edge modes. As such, we have
promoted the periodic table of integer topological
phases to its fractional counterpart.
The following paradigms were applied.\\
(1) Each Luttinger liquid wire describes (spinfull or spinless) electrons. 
We rely on a bosonized description.\\
(2) Back-scattering and short-range interactions within and between wires 
are added. Modes are gapped out if these terms acquire 
a finite expectation value.\\
(3) A mutual compatibility condition, 
the Haldane criterion, 
is imposed among the terms that acquire an expectation value. 
It is an incarnation of the statement that the operators have to commute
if they are to be replaced simultaneously by their expectation values.\\
(4) A set of discrete and local symmetries are imposed on all terms 
in the Hamiltonian. When modes become massive, 
they may not break these symmetries.\\
(5) We do not study the renormalization group flow of the interaction and 
back-scattering terms, but analyze the model in a strong-coupling limit. 

Using this strategy, the following directions present themselves for
future work. 

First, for symmetry class A, 
we have shown that sufficiently strong interactions among identical electrons 
can turn any topological phase with the same topological number 
controlling both the Hall and thermal conductivities 
into a SRE topological phase with independent quantized values 
of the Hall and thermal conductivities. 
[We only need to make the interaction encoded by Eq.%
~(\ref{eq: definition of interactions that produce bosons out of fermions})
dominant.]
Hence, it is natural to seek a putative breakdown 
of the topological counterpart to the Wiedemann-Franz law for metals
in the symmetry class AII and for the LRE phases in the symmetry
classes A and AII.

Second, we can impose on our wire construction additional, albeit less generic,
symmetries such as a non-local inversion symmetry or
such as a residual $U(1)$ spin symmetry.

Third, our construction can be extended to
topological phases of systems that have bosons as their elementary
degrees of freedom. For bosons, no analogue of the tenfold way exists
to provide guidance. However, several works are dedicated to the
classification of SRE and LRE phases of bosons, which might provide a
helpful starting point.%
~\cite{Lu12}

Fourth, extensions to higher dimensions could be considered.~\cite{Senthil1,Senthil2}  This
would, however, entail leaving the comfort zone of one-dimensional
bosonization, with a necessary generalization of the Haldane criterion in
a layer construction.


\section*{Acknowledgments}
This work was supported by the European Research Council through 
the grant TOPOLECTRICS, ERC-StG-Thomale-336012, by DARPA SPAWARSYSCEN Pacific N66001-11-1-4110 (T.N.),  and by DOE Grant DEF-06ER46316 (C.C.).

\appendix

\begin{widetext}

\section{Conditions for particle-hole and time-reversal symmetry}
\label{appendixsec: cond TRS and PHS}

The conditions%
~(\ref{eq: TRS H})
and~(\ref{eq: PHS H}) for TRS and PHS
can be derived by adapting
the derivations
\begin{subequations}
\begin{equation}
\begin{split}
\widehat{\Theta}\,
\hat{H}^{\,}_{\{\mathcal{T}\}}\,
\widehat{\Theta}^{-1}=&\,
\int\mathrm{d}x
\sum_{\mathcal{T}}
h^{\,}_{\mathcal{T}}
\cos
\left(
\mathcal{T}^{\mathsf{T}}\,
\mathcal{K}(\mathcal{P}^{\,}_{\Theta}\,
\widehat{\Phi}
+
\pi\, 
\mathcal{K}^{-1}\,
\mathcal{I}^{\,}_{\Theta})
+
\alpha^{\,}_{\mathcal{T}}
\right)
\\
=&\,
\int\mathrm{d}x
\sum_{\mathcal{T}}
h^{\,}_{-\mathcal{P}^{\,}_{\Theta}\,\mathcal{T}}\,
\cos
\left(
\mathcal{T}^{\mathsf{T}}\, 
\mathcal{K}\, 
\widehat{\Phi}
-
\pi\,
\mathcal{T}^{\mathsf{T}}\, 
\mathcal{P}^{\,}_{\Theta}\,
\mathcal{I}^{\,}_{\Theta}
+
\alpha^{\,}_{-\mathcal{P}_{\Theta}\,\mathcal{T}}
\right)
\\
\stackrel{!}{=}&\,
\int\mathrm{d}x
\sum_{\mathcal{T}}
h^{\,}_{\mathcal{T}}
\cos
\left(
\mathcal{T}^{\mathsf{T}}\, 
\mathcal{K}\, 
\widehat{\Phi}
+
\alpha^{\,}_{\mathcal{T}}
\right)
\end{split}
\end{equation}
and
\begin{equation}
\begin{split}
\widehat{\Pi}\,
\hat{H}^{\,}_{\{\mathcal{T}\}}\,
\widehat{\Pi}^{-1}=&\,
\int\mathrm{d}x
\sum_{\mathcal{T}}
h^{\,}_{\mathcal{T}}
\cos
\left(
\mathcal{T}^{\mathsf{T}}\,
\mathcal{K}(\mathcal{P}^{\,}_{\Pi}\,
\widehat{\Phi}
+
\pi\, 
\mathcal{K}^{-1}\,
\mathcal{I}^{\,}_{\Pi})
+
\alpha^{\,}_{\mathcal{T}}
\right)
\\
=&\,
\int\mathrm{d}x
\sum_{\mathcal{T}}
h^{\,}_{\mathcal{P}^{\,}_{\Pi}\,\mathcal{T}}\,
\cos
\left(
\mathcal{T}^{\mathsf{T}}\, 
\mathcal{K}\, 
\widehat{\Phi}
+
\pi\,
\mathcal{T}^{\mathsf{T}}\, 
\mathcal{P}^{\,}_{\Pi}\,
\mathcal{I}^{\,}_{\Pi}
+
\alpha^{\,}_{\mathcal{P}_{\Pi}\,\mathcal{T}}
\right)
\\
\stackrel{!}{=}&\,
\int\mathrm{d}x
\sum_{\mathcal{T}}
h^{\,}_{\mathcal{T}}
\cos
\left(
\mathcal{T}^{\mathsf{T}}\, 
\mathcal{K}\, 
\widehat{\Phi}
+
\alpha^{\,}_{\mathcal{T}}
\right)
\end{split}
\label{eq: condition imposed by Pi on H T}
\end{equation} 
\end{subequations}
of Eqs.~(\ref{eq: TRS H d})
and~(\ref{eq: TRS H e}),
respectively.

\section{Symmetry class C}
\label{appendixsec: Symmetry class C}

Class C is defined 
on line 9 of Table~\ref{table: main table}
by the operator 
$\widehat{\Pi}$
for the PHS obeying 
$\Pi^{2}=-1$
with neither TRS nor chiral symmetry present
(as is implied by the entries 0 for $\Theta^{2}$
and $C^{2}$ in Table \ref{table: main table}). 
In physical terms, class C describes a generic superconductor
for which full spin $SU(2)$ symmetry is retained but TRS is broken.
The only difference to the case of class D considered 
in the main text is that the number of degrees of freedom is doubled. 
We postulate that under PHS the following transformation rules hold 
\begin{equation}
b^{\dag}_{\uparrow,\mathrm{R}}\stackrel{\widehat{\Pi}}{\rightarrow}
-b^{\,}_{\downarrow,\mathrm{R}},
\qquad
b^\dagger_{\downarrow,\mathrm{R}}\stackrel{\widehat{\Pi}}{\rightarrow}
+b^{\,}_{\uparrow,\mathrm{R}},
\end{equation}
for the creation operators of Bogoliubov-deGennes quasiparticles
that are right (R) movers at the Fermi energy and
carry the spin quantum numbers $\uparrow,\downarrow$.
We apply the same transformation law to 
the creation operators of Bogoliubov-deGennes quasiparticles
that are left (L) movers at the Fermi energy and
carry the spin quantum numbers $\uparrow,\downarrow$.

We consider identical wires with quasiparticles of type 1 and 2,
``particles'' and ``holes'', as well as left- and right-moving 
degrees of freedom. For any given wire with the basis 
(
$b^{\dag}_{\uparrow,\mathrm{L}}$, 
$b^{\dag}_{\downarrow,\mathrm{L}}$,
$b^{\dag}_{\uparrow,\mathrm{R}}$, 
$b^{\dag}_{\downarrow,\mathrm{R}}$,
$b^{\,}_{\uparrow,\mathrm{R}}$, 
$b^{\,}_{\downarrow,\mathrm{R}}$,
$b^{\,}_{\uparrow,\mathrm{L}}$, 
$b^{\,}_{\downarrow,\mathrm{L}}$
),
the $K$-matrix reads
\begin{subequations}
\begin{equation}
K:=
\mathrm{diag}\,
(+1,+1,-1,-1,-1,-1,+1,+1),
\label{eq: K class C}
\end{equation}
where PHS has the representation
\begin{equation}
P_{\Pi}:=
\begin{pmatrix}
0&0&0&0&0&0&0&1\\
0&0&0&0&0&0&1&0\\
0&0&0&0&0&1&0&0\\
0&0&0&0&1&0&0&0\\
0&0&0&1&0&0&0&0\\
0&0&1&0&0&0&0&0\\
0&1&0&0&0&0&0&0\\
1&0&0&0&0&0&0&0
\end{pmatrix},
\qquad
I_{\Pi}
:=
\begin{pmatrix}0\\1\\0\\1\\0\\1\\0\\1\end{pmatrix}.
\end{equation}
\end{subequations}

\subsection{SRE phase}

To complete the definition of an array of quantum wires realizing 
a SRE phase in the symmetry class C, we specify the $(4N-2)$
scattering vectors
\begin{subequations}
\begin{align}
\mathcal{T}^{(j)}_{1,\mathrm{SRE}}:=&
(0,0,0,0,0,0,0,0|\cdots|+1,0,-1,0,-1,0,+1,0|
\cdots|0,0,0,0,0,0,0,0),\\
\mathcal{T}^{(j)}_{2,\mathrm{SRE}}:=&
(0,0,0,0,0,0,0,0|\cdots|0,+1,0,-1,0,-1,0,+1|
\cdots|0,0,0,0,0,0,0,0),\\
\mathcal{T}^{(l)}_{3,\mathrm{SRE}}:=&
(0,0,0,0,0,0,0,0|\cdots|0,0,+1,0,-1,0,0,0|0,-1,0,0,0,0,0,+1|
\cdots|0,0,0,0,0,0,0,0),\\
\mathcal{T}^{(l)}_{4,\mathrm{SRE}}:=&
(0,0,0,0,0,0,0,0|\cdots|0,0,0,-1,0,+1,0,0|+1,0,0,0,0,0,-1,0|
\cdots|0,0,0,0,0,0,0,0),
\end{align}
\label{eq: tunneling vectors for C SRE}
\end{subequations}
for $j=1,\cdots,N$, and $l=1,\cdots,N-1$. These scattering vectors gap
out all modes in the bulk and comply both with PHS and with the
Haldane criterion~(\ref{eq: Haldane conditions}).  Of the remaining
four modes, two are localized at either edge of the system. The
remaining two modes on either edge share the same chirality, for they
could be gapped out by the non-local scattering vectors
\begin{subequations}
\begin{align}
\mathcal{T}^{(0)}_{3,\mathrm{SRE}}:=&
(0,-1,0,0,0,0,0,+1|\cdots|0,0,+1,0,-1,0,0,0),\\
\mathcal{T}^{(0)}_{4,\mathrm{SRE}}:=&
(+1,0,0,0,0,0,-1,0|\cdots|0,0,0,-1,0,+1,0,0),
\end{align}
\end{subequations}
which act on modes with $+$ chirality on the left and and of $-$
chirality on the right edge only.  We conclude that the pair of chiral
modes on either edge is protected from backscattering.  Extending
this construction to any integer number of layers yields the
$\mathbb{Z}$ classification of class C.

\subsection{LRE phase}

To complete the definition of an array of quantum wires realizing
a LRE phase in the symmetry class C, we use the $2N$ scattering
vectors $\mathcal{T}^{(j)}_{1,\mathrm{SRE}}$ and
$\mathcal{T}^{(j)}_{2,\mathrm{SRE}}$, $j=1,\cdots, N$ defined in
Eq.~\eqref{eq: tunneling vectors for C SRE} and supplement them with
the $2(N-1)$ scattering vectors
\begin{subequations}
\begin{align}
\mathcal{T}^{(j)}_{3,\mathrm{LRE}}:=&\,
(0,0,0,0,0,0,0,0|\cdots|
0,-m^{\,}_{-},m^{\,}_{+},0,-m^{\,}_{+},0,0,m^{\,}_{-}|0,-m^{\,}_{+},m^{\,}_{-},0,-m^{\,}_{-},0,0,m^{\,}_{+}
|\cdots|0,0,0,0,0,0,0,0),
\\
\mathcal{T}^{(j)}_{4,\mathrm{LRE}}:=&\,
(0,0,0,0,0,0,0,0|\cdots|
m^{\,}_{-},0,0,-m^{\,}_{+},0,m^{\,}_{+},-m^{\,}_{-},0|m^{\,}_{+},0,0,-m^{\,}_{-},0,m^{\,}_{-},-m^{\,}_{+},0
|\cdots|0,0,0,0,0,0,0,0),
\end{align}
\end{subequations}
for $j=1,\cdots, N-1$, and $m^{\,}_{\pm}=(m\pm1)/2$ as well as $m$ an
odd positive integer. These tunneling vectors gap out all modes in the
bulk and comply both with PHS and with the Haldane criterion%
~(\ref{eq: Haldane conditions}). One verifies that there exists a linear
transformation with integer entires $W$ and $|\mathrm{det}\,W|=m^{8}$ such
that
\begin{equation}
\mathcal{T}^{(j)}_{l,\mathrm{SRE}}=
\mathcal{W}^{-1}\,\mathcal{T}^{(j)}_{l,\mathrm{LRE}},
\qquad j=1,\cdots,N-1,
\qquad l=3,4.
\end{equation}
The $K$-matrix transforms according to Eq.~\eqref{eq: K tilde FQHE},
leaving the effective effective edge theory with two chiral modes of
PHS symmetric superpositions of Laughlin quasiparticles with Laughlin
quasiholes on either edge of the system.  As with the SRE phase of
symmetry class C, this is a completely chiral theory and no
back-scattering mechanism can gap out modes by the Haldane
criterion~(\ref{eq: Haldane conditions}).  Extending this construction
to any integer number of layers yields a $\mathbb{Z}$ classification
of the LRE phase in symmetry class C for every positive odd integer
$m$.

\section{Symmetry class DIII}
\label{appendixsec: Symmetry class DIII}

Class DIII is defined 
on line 4 of Table~\ref{table: main table}
by the operator 
$\widehat{\Pi}$
for the PHS obeying 
$\Pi^{2}=+1$ and
with the TRS $\Theta^{2}=-1$. 
In physical terms, class DIII describes a generic superconductor
for which full spin $SU(2)$ symmetry is broken but TRS is retained.
We use the same basis and $K$-matrix as specified for class C 
in Eq.~\eqref{eq: K class C}, namely
\begin{subequations}
\begin{equation}
K:=
\mathrm{diag}\,
(+1,+1,-1,-1,-1,-1,+1,+1).
\label{eq: K class DIII}
\end{equation}
The PHS now has the representation
\begin{equation}
P^{\,}_{\Pi}:=
\begin{pmatrix}
0&0&0&0&0&0&1&0\\
0&0&0&0&0&0&0&1\\
0&0&0&0&1&0&0&0\\
0&0&0&0&0&1&0&0\\
0&0&1&0&0&0&0&0\\
0&0&0&1&0&0&0&0\\
1&0&0&0&0&0&0&0\\
0&1&0&0&0&0&0&0
\end{pmatrix},
\qquad
I^{\,}_{\Pi}:=
\begin{pmatrix}0\\0\\0\\0\\0\\0\\0\\0\end{pmatrix},
\end{equation}
while TRS is defined by
\begin{equation}
P^{\,}_{\Theta}:=
\begin{pmatrix}
0&0&0&1&0&0&0&0\\
0&0&1&0&0&0&0&0\\
0&1&0&0&0&0&0&0\\
1&0&0&0&0&0&0&0\\
0&0&0&0&0&0&0&1\\
0&0&0&0&0&0&1&0\\
0&0&0&0&0&1&0&0\\
0&0&0&0&1&0&0&0
\end{pmatrix},
\qquad
I^{\,}_{\Theta}
:=
\begin{pmatrix}0\\1\\0\\1\\0\\1\\0\\1\end{pmatrix}.
\end{equation}
\end{subequations}

\subsection{SRE phase}

To complete the definition of an array of quantum wires realizing 
a SRE phase in the symmetry class DIII, we specify the $(4N-2)$
tunneling vectors
\begin{subequations}
\begin{align}
\mathcal{T}^{(j)}_{1,\mathrm{SRE}}:=&\,
(0,0,0,0,0,0,0,0|\cdots|+1,0,-1,0,-1,0,+1,0|\cdots|0,0,0,0,0,0,0,0),\\
\mathcal{T}^{(j)}_{2,\mathrm{SRE}}:=&\,
(0,0,0,0,0,0,0,0|\cdots|0,+1,0,-1,0,-1,0,+1|\cdots|0,0,0,0,0,0,0,0),\\
\mathcal{T}^{(l)}_{3,\mathrm{SRE}}:=&\,
(0,0,0,0,0,0,0,0|\cdots|0,0,+1,0,-1,0,0,0|-1,0,0,0,0,0,+1,0|\cdots|0,0,0,0,0,0,0,0),\\
\mathcal{T}^{(l)}_{4,\mathrm{SRE}}:=&\,
(0,0,0,0,0,0,0,0|\cdots|0,-1,0,0,0,0,0,+1|0,0,0,+1,0,-1,0,0|\cdots|0,0,0,0,0,0,0,0),
\end{align}
\label{eq: tunneling vectors for DIII SRE}
\end{subequations}
for $j=1,\cdots,N$, and $l=1,\cdots,N-1$.
These tunnelings gap out all the bulk modes.  
Here, $\mathcal{T}^{(j)}_{1,\mathrm{SRE}}$ and 
$\mathcal{T}^{(j)}_{2,\mathrm{SRE}}$ as
well as $\mathcal{T}^{(j)}_{3,\mathrm{SRE}}$ and 
$\mathcal{T}^{(j)}_{4,\mathrm{SRE}}$ are pairwise
related by TRS, while each of the tunneling vectors is in itself PHS.
The phases of the corresponding cosine terms in Hamiltonian%
~\eqref{eq: hat H bosonized c} 
comply with both TRS and PHS as if
\begin{equation}
\alpha^{\,}_{\mathcal{T}^{(j)}_{1,\mathrm{SRE}}}
=
\alpha^{\,}_{\mathcal{T}^{(j)}_{2,\mathrm{SRE}}},
\qquad
\alpha^{\,}_{\mathcal{T}^{(j)}_{3,\mathrm{SRE}}}
=
\alpha^{\,}_{\mathcal{T}^{(j)}_{4,\mathrm{SRE}}}=0.
\end{equation}

On both wire $j=1$ and wire $j=N$, there remains 
a single Kramers degenerate pair of propagating modes. 
Now,  the tunneling vector
\begin{equation}
\mathcal{T}^{\mathrm{L}}=
(-1,0,0,+1,0,-1,+1,0|\cdots|0,0,0,0,0,0,0,0)
\end{equation} 
acts locally on the left edge, satisfies the Haldane
criterion with all existing scattering vectors,
and is unique up to an integer multiplicative factor.
It might thus be concluded that the left pair of Kramers degenerate
edge states can be gapped by the tunneling $\mathcal{T}^{\mathrm{L}}$.
This is not so however. Indeed,
while $\mathcal{T}^{\mathrm{L}}$ itself is both compliant with PHS and TRS,
its contribution to $\hat{H}^{\,}_{\{\mathcal{T}\}}$ induces
another expectation value that breaks TRS spontaneously. 
To see this, we note that, by the particle-hole redundancy, the bosonic
fields $\hat{\phi}^{\,}_{1}(x)$ and $-\hat{\phi}^{\,}_{7}(x)$ as well as
$\hat{\phi}^{\,}_{4}(x)$ and $-\hat{\phi}^{\,}_{6}(x)$ have to be identified. 
Thus, it is $\cos\big(T^{\mathrm{L}\,\mathsf{T}}\,K\,\hat{\Phi}(x)\big)\sim
\cos\big(2\hat{\phi}^{\,}_{1}(x)-2\hat{\phi}^{\,}_{4}(x)\big)$
that acquires an expectation value. Now,
the term 
$\cos\big(\hat{\phi}^{\,}_{1}(x)-\hat{\phi}^{\,}_{4}(x)\big)$
is more relevant from the point of view of the renormalization group
than 
$\cos\big(2\hat{\phi}^{\,}_{1}(x)-2\hat{\phi}^{\,}_{4}(x)\big)$.
If
$\cos\big(2\hat{\phi}^{\,}_{1}(x)-2\hat{\phi}^{\,}_{4}(x)\big)$
acquires an expectation value, so does
$\cos\big(\hat{\phi}^{\,}_{1}(x)-\hat{\phi}^{\,}_{4}(x)\big)$.
However, 
$\cos\big(\hat{\phi}^{\,}_{1}(x)-\hat{\phi}^{\,}_{4}(x)\big)$
corresponds to
\begin{equation}
\bar{\mathcal{T}}^{\mathrm{L}}=
(-1,0,0,+1,0,0,0,0|\cdots|0,0,0,0,0,0,0,0)
\end{equation}
(and scattering vectors related by PHS) must then break TRS spontaneously, 
for the resulting condition 
$\alpha^{\,}_{\bar{\mathcal{T}}^{\mathrm{L}}}=
\alpha^{\,}_{-\mathcal{P}_{\Theta}\bar{\mathcal{T}}^{\mathrm{L}}}+\pi=
\alpha^{\,}_{\bar{\mathcal{T}}^{\mathrm{L}}}+\pi$ 
on the phase of its cosine term cannot be met.
If we rule out the spontaneous breaking of TRS on the left edge,
we must rule out the tunnelings $n\,\mathcal{T}^{\mathrm{L}}$ for any integer
$n$. Under this assumption, there remains a single gapless
left pair of Kramers degenerate edge states. 

We conclude that there is no possibility to localize the remaining
edge modes with perturbations that comply with both TRS and PHS.  Had
we considered two layers of this wire model, edge modes in both layers
can be gapped out pairwise, similar to the case of class AII that we
discussed in the main text. We conclude that the SRE phase of symmetry
class DIII features a $\mathbb{Z}^{\,}_{2}$ topological classification.

\subsection{LRE phase}

To complete the definition of an array of quantum wires realizing
a LRE phase in the symmetry class D III, we use the $2N$ scattering
vectors $\mathcal{T}^{(j)}_{1,\mathrm{SRE}}$ and
$\mathcal{T}^{(j)}_{2,\mathrm{SRE}}$, $j=1,\cdots, N$ defined in
Eq.~\eqref{eq: tunneling vectors for DIII SRE} and supplement them
with the $2(N-1)$ scattering vectors
\begin{subequations}
\begin{align}
\mathcal{T}^{(j)}_{3,\mathrm{LRE}}:=&\,
(0,0,0,0,0,0,0,0|\cdots|
-m^{\,}_{-},0,m^{\,}_{+},0,-m^{\,}_{+},0,m^{\,}_{-},0|-m^{\,}_{+},0,m^{\,}_{-},0,-m^{\,}_{-},0,m^{\,}_{+},0
|\cdots|0,0,0,0,0,0,0,0),
\\
\mathcal{T}^{(j)}_{4,\mathrm{LRE}}:=&\,
(0,0,0,0,0,0,0,0|\cdots|
0,-m^{\,}_{+},0,m^{\,}_{-},0,m^{\,}_{-},0,m^{\,}_{+}|0,-m^{\,}_{-},0,m^{\,}_{+},0,-m^{\,}_{+},0,m^{\,}_{-}
|\cdots|0,0,0,0,0,0,0,0),
\end{align}
\end{subequations}
for $j=1,\cdots, N-1$, and $m^{\,}_{\pm}=(m\pm1)/2$ as well as $m$ an
odd positive integer. These tunneling vectors gap out all modes in the
bulk and comply both with PHS and with the Haldane criterion%
~(\ref{eq: Haldane conditions}). One verifies that there exists a linear
transformation with integer entires $W$ and $|\mathrm{det}\,W|=m^{8}$ such
that
\begin{equation}
\mathcal{T}^{(j)}_{l,\mathrm{SRE}}=
\mathcal{W}^{-1}\,
\mathcal{T}^{(j)}_{l,\mathrm{LRE}},
\qquad j=1,\cdots,N-1,
\qquad l=3,4.
\end{equation}
The $K$-matrix transforms according to Eq.~\eqref{eq: K tilde FQHE},
leaving the effective effective edge theory with one Kramers degenerate pair of
PHS symmetric superpositions of Laughlin quasiparticles with Laughlin
quasi-holes on either edge of the system.  As with the SRE phase of
symmetry class DIII, this edge theory is protected by PHS and TRS. Two
copies of it, however, can be fully gapped out while preserving PHS
and TRS.  This yields a $\mathbb{Z}^{\,}_{2}$ classification of the
LRE phase in symmetry class DIII for every positive odd integer $m$.

\end{widetext}


\begin{thebibliography}{10}

\bibitem{Kane05a}
C.~L. Kane and E.~J. Mele, Phys. Rev. Lett. {\bf 95},  146802  (2005).

\bibitem{Kane05b}
C.~L. Kane and E.~J. Mele, Phys. Rev. Lett. {\bf 95},  226801  (2005).

\bibitem{Bernevig06}
B.~A. Bernevig, T.~L. Hughes, and S.-C. Zhang, Science {\bf 314},  1757
  (2006).

\bibitem{Koenig07}
M. K\"onig, S. Wiedmann, C. Br\"une, A. Roth, H. Buhmann, L.~W. Molenkamp,
  X.-L. Qi, and S.-C. Zhang, Science {\bf 318},  766  (2007).

\bibitem{Klitzing80}
K.~v. Klitzing, G. Dorda, and M. Pepper, Phys. Rev. Lett. {\bf 45},  494
  (1980).

\bibitem{Laughlin81}
R.~B. Laughlin, Phys. Rev. B {\bf 23},  5632  (1981).

\bibitem{Thouless82}
D.~J. Thouless, M. Kohmoto, M.~P. Nightingale, and M. den Nijs, Phys. Rev.
  Lett. {\bf 49},  405  (1982).

\bibitem{Kane07}
L. Fu, C.~L. Kane, and E.~J. Mele, Phys. Rev. Lett. {\bf 98},  106803  (2007).

\bibitem{Schnyder08}
A.~P. Schnyder, S. Ryu, A. Furusaki, and A.~W.~W. Ludwig, Phys. Rev. B {\bf
  78},  195125  (2008).

\bibitem{Kitaev09}
A. Kitaev, AIP Conf. Proc. {\bf 1134},  22  (2009).

\bibitem{Gurarie11}
V. Gurarie, Phys. Rev. B {\bf 83},  085426  (2011).

\bibitem{Fidkowski10}
L. Fidkowski and A. Kitaev, Phys. Rev. B {\bf 81},  134509  (2010).

\bibitem{Manmana12}
S.~R. Manmana, A.~M. Essin, R.~M. Noack, and V. Gurarie, Phys. Rev. B {\bf 86},
   205119  (2012).

\bibitem{Wang12}
Z. Wang and S.-C. Zhang, Phys. Rev. X {\bf 2},  031008  (2012).

\bibitem{Raghu08}
S. Raghu, X.-L. Qi, C. Honerkamp, and S.-C. Zhang, Phys. Rev. Lett. {\bf 100},
  156401  (2008).

\bibitem{Budich12}
J.~C. Budich, R. Thomale, G. Li, M. Laubach, and S.-C. Zhang, Phys. Rev. B {\bf
  86},  201407  (2012).

\bibitem{Gu09}
Z.-C. Gu and X.-G. Wen, Phys. Rev. B {\bf 80},  155131  (2009).

\bibitem{Pollmann10}
F. Pollmann, A.~M. Turner, E. Berg, and M. Oshikawa, Phys. Rev. B {\bf 81},
  064439  (2010).

\bibitem{Cirac11}
N. Schuch, D. Perez-Garcia, and I. Cirac, Phys. Rev. B {\bf 84},  165139
  (2011).

\bibitem{Chen11a}
X. Chen, Z.-C. Gu, and X.-G. Wen, Phys. Rev. B {\bf 83},  035107  (2011).

\bibitem{Chen11b}
X. Chen, Z.-C. Gu, and X.-G. Wen, Phys. Rev. B {\bf 84},  235128  (2011).

\bibitem{Fidkowski11}
L. Fidkowski and A. Kitaev, Phys. Rev. B {\bf 83},  075103  (2011).

\bibitem{Turner11}
A.~M. Turner, F. Pollmann, and E. Berg, Phys. Rev. B {\bf 83},  075102  (2011).

\bibitem{Gu12}
Z.-C. {Gu} and X.-G. {Wen}, ArXiv e-prints  (2012).

\bibitem{Lu12}
Y.-M. Lu and A. Vishwanath, Phys. Rev. B {\bf 86},  125119  (2012).

\bibitem{Chen13}
X. Chen, Z.-C. Gu, Z.-X. Liu, and X.-G. Wen, Phys. Rev. B {\bf 87},  155114
  (2013).

\bibitem{Lu13}
Y.-M. {Lu} and A. {Vishwanath}, ArXiv e-prints  (2013).

\bibitem{Wen91}
X.-G. Wen, Int. J. Mod. Phys. {\bf B5},  1641  (1991).

\bibitem{Yakovenko91}
V.~M. Yakovenko, Phys. Rev. B {\bf 43},  11353  (1991).

\bibitem{Lee94}
D.-H. Lee, Phys. Rev. B {\bf 50},  10788  (1994).

\bibitem{Kane02}
C.~L. Kane, R. Mukhopadhyay, and T.~C. Lubensky, Phys. Rev. Lett. {\bf 88},
  036401  (2002).

\bibitem{Teo14}
J.~C.~Y. Teo and C.~L. Kane, Phys. Rev. B {\bf 89},  085101  (2014).

\bibitem{Sondhi01}
S.~L. Sondhi and K. Yang, Phys. Rev. B {\bf 63},  054430  (2001).

\bibitem{Klinovaja13b}
J. Klinovaja and D. Loss, Phys. Rev. Lett. {\bf 111},  196401  (2013).

\bibitem{Klinovaja13a}
J. Klinovaja and D. Loss, arxiv 1305.1569  .

\bibitem{gangof11}
R.~S.~K. Mong {\it et~al.}, Phys. Rev. X {\bf 4},  011036  (2014).

\bibitem{Seroussi14}
I. Seroussi, E. Berg, and Y. Oreg,   (2014).

\bibitem{PhysRevX.4.031009}
A. Vaezi, Phys. Rev. X {\bf 4},  031009  (2014).

\bibitem{Haldane95}
F.~D.~M. Haldane, Phys. Rev. Lett. {\bf 74},  2090  (1995).

\bibitem{Wen91a}
X.~G. Wen, Phys. Rev. B {\bf 44},  2664  (1991).

\bibitem{Sachdev91}
S. Sachdev and N. Read, Int. J. Mod. Phys. B {\bf 05},  219  (1991).

\bibitem{Mudry94}
C. Mudry and E. Fradkin, Phys. Rev. B {\bf 49},  5200  (1994).

\bibitem{footnotechiraltrsf}
A chiral symmetry is preseent if there exists a chiral operator $\widehat{C}$
  that is antiunitary and commutes with the Hamiltonian. The single-particle
  representation $C$ of $\widehat{C}$ is a unitary operator that anticommutes
  with the single-particle Hamiltonian. In a basis in which $C$ is strictly
  block off diagonal, $C$ reverses the chirality. This chirality is unrelated
  to the direction of propagation of left and right movers which is also called
  chirality in this paper.

\bibitem{Neupert11}
T. Neupert, L. Santos, S. Ryu, C. Chamon, and C. Mudry, Phys. Rev. B {\bf 84},
  165107  (2011).

\bibitem{Altland97}
A. Altland and M.~R. Zirnbauer, Phys. Rev. B {\bf 55},  1142  (1997).

\bibitem{CJNSP_Majorana_fields}
G. Goldstein and C. Chamon, Phys. Rev. B {\bf 86},  115122  (2012).

\bibitem{Haldane88}
F.~D.~M. Haldane, Phys. Rev. Lett. {\bf 61},  2015  (1988).

\bibitem{footnote:stabilityAII}
Even integer multiples of $\mathcal{T}^{\,}_{\mathrm{eff}}$ would gap the edge
  states, but they must also be discarded as explained in
  Ref.~\onlinecite{Neupert11}.

\bibitem{Laughlin83}
R.~B. Laughlin, Phys. Rev. Lett. {\bf 50},  1395  (1983).

\bibitem{Levin09}
M. Levin and A. Stern, Phys. Rev. Lett. {\bf 103},  196803  (2009).

\bibitem{scharfenberger}
B. Scharfenberger, R. Thomale, and M. Greiter, Phys. Rev. B {\bf 84},  140404
  (2011).

\bibitem{maissam1}
M. Barkeshli and X.-L. Qi, Phys. Rev. X {\bf 2},  031013  (2012).

\bibitem{maissam2}
M. Barkeshli, C.-M. Jian, and X.-L. Qi, Phys. Rev. B {\bf 87},  045130  (2013).

\bibitem{maissam3}
M. Barkeshli, C.-M. Jian, and X.-L. Qi, Phys. Rev. B {\bf 88},  235103  (2013).

\bibitem{abolhassan}
A. Vaezi, Phys. Rev. B {\bf 87},  035132  (2013).

\bibitem{netanel}
N.~H. Lindner, E. Berg, G. Refael, and A. Stern, Phys. Rev. X {\bf 2},  041002
  (2012).

\bibitem{jason1}
D.~J. Clarke, J. Alicea, and K. Shtengel, Nature Comm. {\bf 4},  1348  (2013).

\bibitem{cheng1}
M. Cheng, Phys. Rev. B {\bf 86},  195126  (2012).

\bibitem{kitaev}
A.~Y. Kitaev, Phys.-Usp. {\bf 44},  131  (2001).

\bibitem{fendley}
P. Fendley, J. Stat. Mech.  P11020  (2012).

\bibitem{nicanded}
N. Read and E. Rezayi, Phys. Rev. B {\bf 59},  8084  (1999).

\bibitem{nacsl1}
M. Greiter and R. Thomale, Phys. Rev. Lett. {\bf 102},  207203  (2009).

\bibitem{nacsl2}
M. Greiter, D.~F. Schroeter, and R. Thomale, Phys. Rev. B {\bf 89},  165125
  (2014).

\bibitem{sela}
Y. Oreg, E. Sela, and A. Stern, Phys. Rev. B {\bf 89},  115402  (2014).

\bibitem{Klinovaja13}
J. {Klinovaja} and D. {Loss}, ArXiv e-prints  (2013).

\bibitem{Senthil1}
C. Wang, A.~C. Potter, and T. Senthil, Science {\bf 343},  629  (2014).

\bibitem{Senthil2}
C. Wang and T. Senthil, Phys. Rev. B {\bf 89},  195124  (2014).

\end{thebibliography}
\end{document}